\documentclass[11pt]{article}

\usepackage{fancyhdr}

\usepackage{geometry}

\RequirePackage[numbers]{natbib}

\newcommand{\prob}{\mathrm{Pr}}

\newcommand\ind{\bot\hspace*{-6pt}\bot}  % {\bot\!\!\!\!\bot}
\newcommand{\bc}{\begin{center}}
	\newcommand{\ec}{\end{center}}

\newcommand{\bit}{\begin{itemize}}
	\newcommand{\eit}{\end{itemize}}

\newcommand{\be}{\begin{eqnarray*}}
	\newcommand{\ee}{\end{eqnarray*}}
\newcommand{\ben}{\begin{eqnarray}}
	\newcommand{\een}{\end{eqnarray}}
\newcommand{\g}{\,\vert\,}

\newcommand{\D}{\mathcal{D}}
\newcommand{\E}{\mathcal{E}}
\newcommand{\G}{\mathcal{G}}

\newcommand{\N}{\mathcal{N}}
\newcommand{\W}{\mathcal{W}}

\newcommand{\pa}{\mathrm{pa}}

\newcommand{\BF}{\mathrm{BF}}

\newcommand{\bzero}{\bm{0}}

\newcommand{\bD}{\bm{D}}

\newcommand{\bL}{\bm{L}}

\newcommand{\bS}{\bm{S}}

\newcommand{\bU}{\bm{U}}
\newcommand{\bX}{\bm{X}}
\newcommand{\bY}{\bm{Y}}
\newcommand{\bZ}{\bm{Z}}

\newcommand{\bx}{\bm{x}}
\newcommand{\by}{\bm{y}}
\newcommand{\bz}{\bm{z}}

\newcommand{\bSigma}{\bm{\Sigma}}

\newcommand{\bOmega}{\bm{\Omega}}

\newcommand{\btheta}{\bm{\theta}}

\usepackage{tikz}
\usetikzlibrary{graphs}
\usetikzlibrary{positioning}

\newcommand{\black}{\color{black}}

\newcommand{\brown}{\color{brown}}

\RequirePackage{amsthm,amsmath,amsfonts,amssymb}
\RequirePackage[numbers]{natbib}
\RequirePackage[colorlinks,citecolor=blue,urlcolor=blue]{hyperref}
\RequirePackage{graphicx}

\usepackage{cancel}

\makeatletter %% <- make @ usable in command names
\newcommand*\Neg[2][0mu]{\Neginternal{#1}{\negslash}{#2}}

\newcommand*\Neginternal[3]{\mathpalette\Neg@{{#1}{#2}{#3}}}
\newcommand*\Neg@[2]{\Neg@@{#1}#2}
\newcommand*\Neg@@[4]{%
	\mathrel{\ooalign{%
			$\m@th#1#4$\cr
			\hidewidth$\m@th#3{#1}\mkern\muexpr#2*2$\hidewidth\cr
	}}%
}

\newcommand*\negslash[1]{\m@th#1\not\mathrel{\phantom{=}}}
\newcommand*\snegslash[1]{\rotatebox[origin=c]{60}{$\m@th#1-$}}
\newcommand*\ssnegslash[1]{\rotatebox[origin=c]{60}{$\m@th#1{\dabar@}\mkern-7mu{\dabar@}$}}
\newcommand*\sssnegslash[1]{\rotatebox[origin=c]{60}{$\m@th#1\dabar@$}}
\makeatother  %% <- revert @

\usepackage{multirow}
\usepackage{amsfonts}
\usepackage{bm}
\usepackage{bbm}
\RequirePackage{amsthm,amsmath,amsfonts,amssymb}
\RequirePackage[numbers]{natbib}
\RequirePackage[colorlinks,citecolor=blue,urlcolor=blue]{hyperref}
\RequirePackage{graphicx}

\usepackage[ruled,vlined,linesnumbered,lined,boxed,commentsnumbered]{algorithm2e}
\SetKwInput{KwInput}{Input}
\SetKwInput{KwOutput}{Output}

\usepackage{tikz}
\usetikzlibrary{arrows.meta}

\usepackage{soul}	
\newtheorem{definition}{Definition}[section]

\theoremstyle{plain}

\RequirePackage[numbers]{natbib}

\usepackage{bm}
\usepackage{bbm}
\usepackage{float}
\usepackage{tikz}
\usepackage{amsthm}
\usepackage{amssymb}
\usepackage{subfig}
\usepackage{longtable}
\usepackage{mathtools,cancel}
\usepackage{url}
\usepackage[ruled,vlined,linesnumbered,lined,boxed,commentsnumbered]{algorithm2e}
\usetikzlibrary{positioning}
\usetikzlibrary{graphs} % LATEX and plain TEX
\usetikzlibrary[graphs]
\usepackage{multirow}
\usepackage{lscape}
\usepackage[english]{babel}
\usepackage{amsmath}
\usepackage{authblk}
\usepackage{xr}

\usepackage{tikz}
\usetikzlibrary{positioning}

\RequirePackage[OT1]{fontenc}
\RequirePackage{amsthm,amsmath}

\newcommand{\baa}{\begin{eqnarray}}
\newcommand{\eaa}{\end{eqnarray}}

\begin{document}

\title{Bayesian sample size determination for causal discovery}
\author[1]{Federico Castelletti \thanks{federico.castelletti@unicatt.it}}
\author[2]{Guido Consonni \thanks{guido.consonni@unicatt.it}}
\affil[1,2]{Department of Statistical Sciences, Universit\`{a} Cattolica del Sacro Cuore, Milan}

%\title{Equivalence class selection of categorical graphical models}
%\author{Federico Castelletti \& Stefano Peluso
%\footnote{Department of Statistical Sciences, Universit\`{a} Cattolica del Sacro Cuore, Milan, federico.castelletti@unicatt.it and  
%Universit\`{a} degli Studi di Milano-Bicocca, Milan, stefano.peluso@unimib.it}}

\date{}

\maketitle

\begin{abstract}

Graphical models based on Directed Acyclic Graphs (DAGs) are widely used to answer causal questions across a variety of scientific and social disciplines. However, observational data alone \black  cannot distinguish in general between DAGs representing the same conditional independence assertions (Markov equivalent DAGs); as a consequence
%\st{, which implies a limited ability to learn causal relationships among variables.}
the orientation of some edges in the graph  remains indeterminate. Interventional data,  produced by exogenous manipulations of  variables in the network,   enhance  the process of  structure learning  because they allow to distinguish among equivalent DAGs, thus sharpening causal inference. Starting from an equivalence class of DAGs, a few  procedures have been devised to produce a collection of variables to be manipulated in order to  identify a causal DAG.
%per He Geng il DAG finale è nella classe di equiv di partenza; non è cosi' per H e Buhl 2014 in qaunto la loro procedura sequenziale determina ad ogni step una nuova classe I-Markov equivalent
Yet, these algorithmic approaches
do not determine the sample size of the interventional data required to   obtain a desired level of statistical accuracy.
%------------------
We tackle this problem from a Bayesian experimental design perspective,
% and adopt the Bayes Factor as a measure of evidence between competing hypotheses.
taking as input a sequence of target variables to be manipulated to
identify edge orientation.
We then propose a method to  determine, at each intervention,  the  optimal sample size
capable of producing a successful experiment  based on a pre-experimental evaluation of the overall probability of substantial correct evidence.

\vspace{0.7cm}

\end{abstract}

\section{Introduction}

%\red For \black citation: insert DOI as well.

\subsection{Causal Directed Acyclic Graphs}

Graphical models based on Directed Acyclic Graphs (DAGs) are widely used to represent
dependence relations among a set of variables; see \citet{Laur:1996} - to which we refer for
graph-theoretic definitions and concepts,
\citet{Cowe:Dawi:Laur:Spie:1999}, \citet{Koller:Friedman:2009}. Applications of DAG models in various
scientific areas abound, especially in genomics; see for instance \citet{Friedman:2004},
\citet{Sachs:etal:2005}, \citet{Shoj:Mich:2009}, \citet{Nagarajan:2013}.
The conditional independencies
expressed by a DAG can be determined using the graphical notion of d-separation \citep{Pear:2000}. Under
faithfulness \citep{Spir:Glym:Sche:2000,Sadeghi:2017},
these independencies are exactly those entailed by the joint distribution
of the variables
which admits a factorization according to the DAG.
%\black see Equation \eqref{eq:DAG:factorization}.
However, it is well known that distinct DAGs can encode the same
set of conditional independencies,
and their collection
is named Markov equivalence class.
Unfortunately, one cannot distinguish between Markov equivalent DAGs using
observational data alone \citep{Chic:2002},
without imposing specific assumptions on the sampling distribution \citep{Peters:Buehlmann:2014}.
For each Markov equivalence class there exists a unique
completed \textit{partially directed} acyclic graph (CPDAG), also named essential graph (EG) \citep{Ande:etal:1997}, which can be taken as representative of the class.
A CPDAG is a special chain graph \citep{Laur:1996}  whose chain
components are \textit{decomposable} undirected graphs (UG) linked by arrowheads.

In practice the structure of a
DAG governing the joint distribution of the observations is unknown, and so is the corresponding CPDAG.
%Although there are fewer CPDAGs than DAGs, their number still increases super-exponentially
%with the number of vertices (Gillispie and Perlman, 2002). For this reason, structural
%learning in the space of CPDAGs has been confined to small graphs.
Learning the structure of a CPDAG has been the subject of several papers.
In the frequentist framework, the two most popular methods are the
Greedy Equivalence Search (GES) of
\citet{Chic:2002} and the PC-algorithm of \citet{Spir:Glym:Sche:2000}, later extended to high-dimensional settings by
\citet{Kalish:Buehlmann:2007}. Specifically, GES is a score-based method which
provides a CPDAG estimate by maximizing a
score function in the space of CPDAGs.
Differently,
the PC algorithm is a constraint based method which outputs an estimate of the true CPDAG using a
sequence of conditional independence tests.
From a Bayesian perspective,
learning a CPDAG is a model selection problem which
can be approached using the Bayes factor \citep{Kass:Raftery:1995} as in \citet{Castelletti:et:al:2018}.
Bayesian inference relies on MCMC methods which explore the space of
Markov equivalence classes and provide an approximate posterior distribution over the space of graphs; see \citet{Madigan:et:al:1996},
\citet{Cast:Perl:2004}, \citet{Sonn:etal:2015} and \citet{He:etal:2013}, who propose a reversible irreducible Markov chain for sparse CPDAGs, having
fewer edges than a small multiple of the number of vertices.

\black

Nowadays DAGs are increasingly used to answer scientific queries in science, technology and society. Typical questions of interest are: \lq \lq which genetic activity is responsible for a particular type of cancer?\rq \rq{}; or \lq \lq what is the effect of introducing a universal basic  income on the level of employment?\rq \rq{}.
%\purple [GC ci sono altre accezioni di causal inference, cfr qui https://arxiv.org/abs/2104.00119 ] \black
If the variables for the problem under consideration can be arranged according to a  DAG structure,    the causal effect
on the response variable due to an external intervention on another variable in the system
can be precisely defined and measured; see \citet{Pear:2000} for a scholarly treatment and  \citet{Pearl:review:Test:2003} for an expository discussion.  \citet{Imbens:2020}
presents a more critical view.
\black
%for a comparative approach between the potential outcome and the DAG approach to causality.

On the other hand,  if all we can learn is
%hand the However, if we can only learn the underlying CPDAG, and  this is not enough for determining a unique causal effect,
a Markov equivalence class, we will obtain a  \emph{collection}  of  causal effects for the same intervention on  a variable (each DAG may potentially produce a distinct value). One strategy  to handle the resulting multiplicities
of effects is to report  lower and upper bounds for the causal effect  conditionally on a selected equivalence class \citep{Maat:etal:2009}.
To reduce this indeterminacy, one could proceed to a Bayesian Model Average (BMA)
of class averages \citep{Castelletti:Consonni:Biom:2021}, where BMA  is with respect to the posterior distribution on the space of equivalence classes.
However sharper
results may be obtained through \emph{interventions},  as we describe in the next subsection.
%\green GC sarei dell'idea di rimuovere il paragrafo che segue in quanto ripete cio' che si dice poco sotto. \purple Rileggendo, mi sembra OK la rimozione
%\black
%\st{Specifically,  once  a promising Markov equivalence class is singled out based on observational data, one can further decompose this class into smaller components  by collecting  further observations  obtained through interventions. This procedure will eventually identify a unique DAG within the equivalence class, as we describe in the next subsection.}

\subsection{DAG identification through interventions}
\label{sec:intro:DAG:identification:interventions}

The starting point for DAG identification  is typically a given CPDAG  which has been estimated based on an initial sample of observational data, and the problem then reduces to orienting the undirected edges in the CPDAG.
%--------
The key idea  to determine edge orientation is to apply  {interventions} on selected nodes (variables) of the graph, i.e. setting exogenously their values.
This can be done   for a single variable,  or jointly for a set of variables, \black  by drawing a value from an external probability distribution, which could also be a point-mass on a pre-determined value. This is called  \emph{perfect} (or hard) \emph{intervention}, and should be contrasted with general (or non-perfect, or soft) intervention \citep{Yang:Uhler:2018}.  In this paper we focus on perfect interventions.

%\purple
The reason why interventions allow to identify the direction of an arrow will become apparent
in Section \ref{sec:BF:edge:orientation}
where we introduce the critical notion of \emph{interventional distribution}.
For the moment suffice it to say that
two \emph{observationally} Markov equivalent DAGs need not be equivalent under interventions, and this fact can be leveraged to split the original equivalence class into smaller interventional equivalence classes. This process can be repeated until each equivalence class contains only a single DAG, so that  identification is achieved. More
on this issue can be found in
\citet{Haus:Buhl:2012,Hauser:Buehlm:2015} who
%\blue
introduce the Greedy Interventional Equivalence Search (GIES) method
as a score-based algorithm for structure learning of interventional equivalence classes and
present several statistical aspects connected to the joint modeling of observational and interventional data.

\black
DAG identification through interventions, also named \emph{active learning}, has been the subject of several contributions over the last two decades or so especially from the computer science community.
%-------------------
\citet{Eber:2008} and \citet{He:etal:2008}
consider the problem of finding interventions that guarantee full identifiability of all DAGs in a given Markov equivalence class which is assumed to be correctly learned.
%---------------
In particular,
\citet{Eber:2008}
proposes a method based on intervention targets of unbounded size, while
% his method was further elaborated in \citet{Hauser:Buehlmann:2014} who prove a result on the number of interventions  for fully identifying a causal model, and provide an algorithm for finding such set of interventions in polynomial time.
%------
% because it will constitute the basis of our method
%for Bayesian sample size determination.
\black
\citet{He:etal:2008} % (H\&G for short)
%is especially relevant for this paper. They
%propose an active learning approach for discovering causal structures
%of DAG models. This method allows to identify the true DAG generating model \emph{via}
%interventions on variables which are selected according to an optimal design strategy which we briefly describe. Given a
%collection of observational data,  a Markov equivalence class is estimated.
%Next the undirected edges in the representative CPDAG are oreinted by manipulating variables
%through external interventions.
deal with single vertex interventions both under  hard and soft interventions.
%, which they call randomized experiments and quasi-experiments, and propose a method for edge orientation of the CPDAG representative of a given equivalence class.
They first show that their method can be implemented locally, that is within each chain component of the CPDAG separately. Next, they propose two kinds of optimal interventional experiments: a \emph{batch} experiment (determining upfront the
minimum set of variables to be manipulated so that undirected edges are all
oriented after the interventions) and
a \emph{sequential}  experiment (start by choosing an intervention variable  such that
the Markov equivalence class can be reduced into a subclass as small
as possible, and then according to the current subclass,  repeatedly select a subsequent variable to be
manipulated until all undirected edges are oriented).
%---------------------
%By intervening on the selected variable, \black they generate a collection of
%interventional data, which is used to
%test  marginal independence between the intervened variable and each of  its clique neighbours using interventional data only.
%%estimate through an independence test
%This leads to a finer collection of interventional Markov equivalence classes.
%Through a suitable sequence of manipulated variables, each
%interventional equivalence class is reduced to a single DAG.
%---------------------
We will return to the approach of \citet{He:etal:2008}  in Section \ref{sec:Bayes:SSD:active:learning}, when we  present  our Bayesian method for sample size determination. \black
%\purple [in realtà diamo dettagli solo sul batch intervention]
\black

%In a simulation study  compare our two active learning approaches to random interventions and an existing approach, and analyze the inﬂuence of estimation errors on  the  overall  performance  of active learning.

\citet{Hauser:Buehlmann:2014} make a significant advancement and propose two methods for active learning  based on sequential intervention experiments. \black
The  first  one  is  a  greedy  approach,  while the second one
yields in polynomial time a minimum set of targets of arbitrary size that guarantees full identifiability.
%\gray
%Qui non è chiaro in che senso il contributo di HB sia significativo: la risposta a mio parere è duplice:
%i) "The key ingredients for the efficiency of OptSingle (Algorithm2) and OptUnb (Algorithm3) are implementations that minimize the objective functions of Eqs.(1) and (2), respectively, without enumerating all DAGs in the equivalence class represented by G"
%ii) l'adozione di un metodo per structure learning (aggiormaneto del CPDAG) che modella congiuntamente obs + int (ma soprattutto i vari int raccolti ad ogni step della procedura sequenziale)
%Scriverei qualcosa tipo: \blue
There are two noteworthy features of their approach.
First, it overcomes some computational inefficiencies related to the enumeration of all DAGs within each chain component
required by \citet{He:etal:2008}.
In addition,  again differently from \citet{He:etal:2008} who implement a testing procedure for edge orientation based  only on the interventional data collected at each  given step, they jointly model all the (observational and) interventional data collected up to that point based on the GIES method \citep{Haus:Buhl:2012}.
As a consequence, the subsequent estimated equivalence class need not  belong to the previous (larger) one, and this can result in a reduction of the estimation error of the whole  active learning procedure. We refer the reader to \citet[Section 5]{Hauser:Buehlmann:2014} for a detailed comparison  of the two approaches and a few others.
Importantly,  their analysis also shows  that the accuracy of each method under investigation  crucially depends on the sample size of the collected data, an important feature which is however investigated only by simulating a few scenarios; see also \citet[Section 7]{Castelletti:Consonni:2020}.
\black
%*********************************************************************
Further relevant papers on active learning are
\citet{Mega:etal:2006},
%also uses single vertex intervention under different utility functions but is not concerned with optimality properties
\citet{Tong;Koll:2001},
\citet{Hytt:etal:JMLR:2013}, and more recently \citet{Klugel:et:al:2019},
%\citet{Castelletti:Consonni:2020},
\citet{squi:etal:arxiv:2020},  \citet{Peng:et:al:2020}.

A feature which is mostly absent in the  works on
active learning
is how many  data to collect  in order to have  \emph{a priori} (i.e. before  data collection)  a reasonable \textit{assurance}  that the adopted method will exhibit desirable inferential properties. In other words,  besides the choice of  variables to intervene upon,  one ought to determine the \emph{sample size} of the interventional data.   This is a typical goal of experimental design,  and one of the objectives  of this paper  is precisely to fill this gap.

\black
\subsection{Bayesian experimental design and sample size determination}
\label{subsec:Bayesian experimental design}
%When deciding upon an experiment several decisions have to be considered. For instance in clinical trials one should decide about which treatment to study, which blocking factors to use, the nature of the units,  how to randomize, the length of time, whether the experiment is obtained in batch or sequentially, the sample size, the proportion of units to allocate to the various treatments.
The Bayesian approach to experimental design has a long tradition.
%------------
%and dates back at least to the second half of the 1950's  as the  paper \citet{Lindley:1956} witnesses. Il paper di Lindley non è proprio adatto; è più una traduzione del metodo di Shannon alla Statistica.
%---------
Lindley was a precursor and supported a decision-theoretic approach; see for instance \citet{lind:1972}.
Following that approach, \citet{Chaloner:Verdinelli:1995} present a unified perspective on the topic with an excellent review up to the mid-1990's. Another almost contemporary review is provided in \citet{Dasgupta:1996}.
%
%\purple io non scriverei molto altro perchè di scarso interesse per questo paper. vado invece subito a SSD

%\subsection{Bayesian sample size determination}
In this paper we focus on a specific aspect of design, namely
Bayesian Sample Size Determination (SSD). This was conceptualized %in a decision theoretic setting
in the influential book \citet{Raiffa:Schlaifer:1961} and has been the subject of several papers in the years to follow.
In the 1997 issue of the \emph{Journal of the Royal Statistical Society Series D} entirely devoted to SSD,  several papers adopted the Bayesian viewpoint, among which we single out
\citet{Lindley:1997} which is  based on the principle of the maximization of expected utility,  \citet{Weiss:1997} which deals with hypothesis testing,
and  \citet{Adcock:1997} which presents a review.
% volume 46, issue 2, of \emph{The Statistician}
%Pham-Gia (1997: è ina discussione di Lindley anche se elaborata), Adcock (1997) [comments on Lindley+, Joseph and Belisle (1997) and Joseph and Wolfson (1997): discuss the merits of using  decision theoretic (e.g. max expected  utility) vs simpler approaches.
%----------------
%\purple verificare se inserire questo The Statistician (44 2, 1995) is also of interest, with papers by Joseph, Wolfson and Du Berger (1995a, b) and discussions by Pham-Gia (1995) and Adcock (1995).:
%\blue In realta'm questo sono paper per SSD nel caos binomiale
%
%\purple questi altri lavori sono citati in Wang and Gelfand (2002): verificare cosa dicono e se sono del filone utility approach o performance approach
%\black   The most recent work is that of Rahme, Joseph, and Gyorkos (2000)
%[è su binomial]
% and Inoue, Berry and Parmigiani (2000)[è un mD Anderson the report; non risulta pubblicato]
%--------
Because of its more pragmatic content,  Bayesian  SSD has been widely analyzed in a variety of applied contexts, notably clinical trials, an early instance being \citet{Spiegelhalter:Freedman:1986}; see also  the comprehensive book by \citet{Spiegelhalter:et:al:book:2003} and references therein.  \citet{O'Hagan:Stevens:2001} carefully distinguished two objectives,  {analysis} and {design},  leading to the use of two distinct  priors for SSD: the \emph{analysis} and the \emph{design} prior. The simultaneous use of two different priors for the same parameter  is actually not new: in a different context it was advocated in an earlier paper by \citet{Etzioni:Kadane:1993}.

Any approach to SSD is predicated on the type of statistical inference one wishes to perform. This is often the test of an hypothesis on a parameter of interest, which typically reduces to  comparing  a simple null hypothesis against a two-sided alternative, or two composite hypotheses, each being one-sided.
\citet[Section 6.5]{Spiegelhalter:et:al:book:2003} discuss   a hybrid, as well as a full,  Bayesian approach to the problem. In the \emph{hybrid} case,  a standard frequentist size-$\alpha$ null-rejection region is considered. Next a prior is assigned to the parameter, and  the classical power function is integrated  with respect to the prior, leading to  an \emph{unconditional}, or  {expected},  \lq \lq classical\rq \rq{} power. Equivalently, one evaluates the (prior)-predictive probability that the test statistic falls in the rejection region of the null hypothesis. Clearly classical \emph{conditional} power used in SSD
%, i.e. the probability of rejecting the null conditionally on a fixed parameter value,
can be recovered as a special case by assigning a degenerate prior  on a fixed  value of the parameter. The optimal sample size is finally derived by requiring that the unconditional power be equal to a pre-specified value,  80\% say.
%------------
The \emph{full} Bayesian approach instead requires first to specify
when the null hypothesis should be rejected,  a sort of \lq \lq Bayesian significance\rq \rq{}. One option is to require that the posterior probability of the null  falls below a fixed threshold.
This probability becomes an event in a pre-posterior analysis, where the observations are yet to be collected, and implicitly defines a rejection region for the null \citep{Spiegelhalter:et:al:book:2003}.

%----------------
If one does not want to use prior probabilities of the hypotheses for SSD, an alternative is  to use the Bayes factor \citep{Kass:Raftery:1995} (BF) directly as a measure of evidence.
This is the approach taken in \citet{Weiss:1997} which considers testing
a point null against a general bilateral alternative under a normal likelihood with known  variance.
A useful feature of this early paper is that it produces the plots of the prior-predictive distribution of the BF under the null and the alternative (represented by a normal prior for the mean parameter). It is apparent that,  for a variety  of reasonable sample sizes, the BF is likely to reach convincing evidence according to traditional scales (e.g. Table \ref{tab:BF:categ}) when the alternative is assumed to hold; while this is hardly the case
when the null is assumed to be true. See \citet{Weiss:1997} for a numerical illustration of this phenomenon.
%Alternatively said, the latter situation would require a much larger sample size.
This \emph{imbalance} in the learning rate happens because the null hypothesis  is nested into the alternative, so that the BF grows essentially as the square root of the sample size under the null, whereas the rate of growth is exponential under the alternative; for a theoretical justification see \citet{Dawid:2011}.
This fact suggests that
treating symmetrically two nested hypothesis for SSD can be problematic.
%For instance in the $N(\theta, \sigma^2=5^2)$ model  with  $\theta \sim N(\mu=2, \tau^2=1)$ setup described in \citet{Weiss:1997} if $H_0: \theta=0 $ and $H_1: \theta \neq 0$, then requiring  $\log BF_{01}> 5$,  corresponding to a $BF_{01}$ exceeding 148, as an acceptance probability of $H_0$,  would imply a sample size  practically impossible to reach if we require the above event to hold with appreciable probability when $H_0$ holds.
%(A $BF_{01}>100$ corresponds to  extreme evidence in favor of $H_0$ according to the modified Jeffreys' scale developed by \citet{Ly:Wagen:et:al:2016}.
%On the other hand, symmetrically accepting $H_1$ if $\log BF_{01}<- 5$ is an event that is much more easily
%achievable under $H_1$.
One possible solution to this problem is setting distinct evidential thresholds for the acceptance of the two hypotheses. An alternative is to  use  a  Bayesian probability of type I error to fix the threshold for rejecting $H_0$, and then determine the sample size required to have a high Bayesian power; these are discussed in \citet{Weiss:1997}.
%; \purple see  also for related ideas Rubin and Stern (1998, Sankhya)[sarebbe da verificare; questo fatto è citato incidentalmente da wang e Gelfand] \black

\citet{Gelfand:Wang:2002}  present a simulation-based framework for Bayesian SSD capable of handling  more complex settings  such as  generalized linear models  and hierarchical models,  as well as planning an experiment for model separation (choice between two models).
Their framework makes a repeated  use of the \emph{fitting} and \emph{sampling} priors, which play the same role of the analysis and design priors of \citet{O'Hagan:Stevens:2001}.
%, whose paper is not referenced, possibly  suggesting an independent development of the concept.

%The work by De Santis and co-authors
\citet{De:Santis:2004} extends the \emph{evidential} approach of \citet{Royal:1997,Royall:2000} to  Bayesian SSD. Since his work introduces important concepts useful also for  this paper, we provide below a short summary.

Consider two hypotheses $H_0$ and $H_1$, and let $y^n$ be a sample of observations  of size $n$. Let $\BF_{01}(y^n)$ be the Bayes factor in favor of $H_0$ against $H_1$, and denote with $p(H_i)$ the prior probability associated to $H_i$, $i=0,1$. For a fixed value $\gamma_0$,  we say that the data provide \emph{decisive} evidence in favor of $H_0$ at level $\gamma_0$ if
$\prob(H_0 \g y^n) >\gamma_0$, equivalently if $\BF_{01}(y^n)>\omega \frac{\gamma_0}{1-\gamma_0}:=k_0$, where $\omega=p(H_0)/p(H_1)$ is the prior odds. Similarly, for a fixed value $\gamma_1$,
the data provide decisive evidence in favor of $H_1$ at level $\gamma_1$ if
$\prob(H_1 \g y^n) >\gamma_1$, equivalently if $\BF_{01}(y^n)<\omega \frac{\gamma_1}{1-\gamma_1}:=1/k_1$.
Once the data come in, the BF will be computed and evaluated against $k_0$ and $k_1$.
For a suitably large value $k_0$,   $\BF_{01}(y^n)>k_0$ will be considered \emph{decisive} evidence in favor of $H_0$, and similarly,  for a large enough $k_1$,
$\BF_{01}(y^n)<1/k_1$  will be considered \emph{decisive} evidence in favor of $H_1$.
While in the exposition so far the values of $k_i$ depend on the threshold probabilities $\gamma_i$ and the prior probabilities $p(H_i)$, one can fix $k_i$  directly having in mind a classification of evidence  based on the BF, such as that provided by
\citet{Schonbrodt:Wagen:2017} which is an adjustment of the original table presented in \citet{Jeffreys:1961}; see Table \ref{tab:BF:categ}.

%\begin{table*}
\begin{table}
	\centering
	\caption{Classification scheme for the interpretation
		of Bayes factor $\textnormal{BF}_{01}$ (from \citet{Schonbrodt:Wagen:2017} adapted from
		\citet{Jeffreys:1961}).}
	\label{tab:BF:categ}
	\begin{tabular}{ccc}
		\hline
		Bayes factor && Evidence category \\
		\hline
		$> 100$ && Extreme evidence for $H_0$ \\
		$30 - 100$ && Very strong evidence for $H_0$ \\
		$10 - 30$ && Strong evidence for $H_0$ \\
		$3 - 10$ && Moderate evidence for $H_0$ \\
		$1 - 3$ && Anecdotal evidence for $H_0$ \\
		$1$ && No evidence \\
		$1/3$ - 1 && Anecdotal evidence for $H_1$ \\
		$1/10$ - 1/3 && Moderate evidence for $H_1$ \\
		$1/30$ - 1/10 && Strong evidence for $H_1$ \\
		$1/100$ - 1/30 && Very strong evidence for $H_1$ \\
		$< 1/100$ && Extreme evidence for $H_1$ \\
		\hline
	\end{tabular}
\end{table}
%\end{table*}
%--------------------
\black
%In particular, $k_0>100$ would provide extreme evidence for $H_0$; $30<k_0<100$ very strong  evidence for $H_0$;
%$10<k_0<30$  strong  evidence for $H_0$, and $3<k_0<10$ only moderate  evidence for $H_0$. Finally $1<k_0<3$ is taken to represent only anecdotal  evidence for $H_0$  Similarly if $k_1<1/100$ we declare
%extreme evidence for $H_1$; $1/100<k_0<1/30$ very strong  evidence for $H_1$, and so on.
%--------------
%It is seen that evidence is categorized into five classes, from extreme down to anecdotal. However to simplify the subsequent exposition
%it is expedient to adopt a binary classification.
Next we declare that $1/k_1<BF_{01}(y^n)<k_0$
corresponds to \emph{inconclusive} evidence, and otherwise \emph{decisive} evidence (either in favor of $H_0$ or $H_1$).
%
%%----------------
%More specifically, \citet{De:Santis:2004}  specifies the following events  whose probability we might wish to control before proceeding to collecting a sample
%\begin{itemize}
%\item
%\emph{Misleading evidence}: %the event that occurs when the
%evidence is decisive in favor
%of the incorrect hypothesis,
%\item
%\emph{Weak evidence}: %the event that occurs when the
%evidence  supports neither
%of the two hypotheses,
%\item
%\emph{Decisive and correct evidence}: %the event that occurs when the
%evidence is decisive
%in favor of the correct hypothesis.
%\end{itemize}

It is instructive to consider the probability
of evidential support provided by the Bayes factor
\emph{conditionally} on each $H_i$.
Thus we obtain
\begin{itemize}
	
	\item
	$p_i^{I}(k_0,k_1,n)$: the probability of \emph{Inconclusive evidence} conditionally on $H_i$,
	
	\item
	$p_i^{DC}(k_i,n)$: the probability of \emph{Decisive and Correct evidence}, namely $\BF_{ij}>k_i$, $i,j=0, 1$ $ i \neq j$,  conditionally on $H_i$,
	
	\item
	$p_i^{M}(k_j,n)=1-p_i^{DC}(k_i,n)-p_i^{I}(k_0,k_1,n)$, the probability of \emph{Misleading evidence} conditionally on $H_i$.
\end{itemize}

Finally one can recover the \emph{unconditional} probability of any of the above types
by averaging the corresponding conditional probability  w.r.t. the prior probabilities $p(H_i)$. In particular we have
\be
\small
p^{DC}(k_0,k_1,n)&=&p(H_0)p_0^{DC}(k_0,n)+p(H_1)p_1^{DC}(k_1,n),
%%\item
%p^{W}(k_0,k_1,n)&=&P(H_0)p_0^{W}(k_0,k_1,n)\\
%&+&
%P(H_1)p_1^{W}(k_0,k_1,n),\\
%%\item
%p^{M}(k_0,k_1,n)&=&P(H_0)p_0^{M}(k_1,n)+P(H_1)p_1^{M}(k_0,n).
\ee
%\begin{itemize}
%\item
%$p^{DC}(k_0,k_1,n)=P(H_0)p_0^{DC}(k_0,n)+P(H_1)p_1^{DC}(k_1,n)$,
%\item
%$p^{W}(k_0,k_1,n)=P(H_0)p_0^{W}(k_0,k_1,n)+P(H_1)p_1^{W}(k_0,k_1,n)$,
%\item
%$p^{M}(k_0,k_1,n)=P(H_0)p_0^{M}(k_1,n)+P(H_1)p_1^{M}(k_0,n)$.
%\end{itemize}
which
represents the overall pre-experimental evaluation of the potential
success of the experiment.
Hence it is proposed to choose the optimal sample size $n^*$
based on $p^{DC}(k_0,k_1,n)$.  Specifically,  for $\zeta \in (0,1)$
\ben
\label{eq:optim:sample:size:general}
n^* = \min\left\{n\in \mathbb{N} : p^{DC}(k_0,k_1,n) \ge \zeta\right\}.
\een
Of course, besides guaranteeing \emph{ex-ante} a fairly high level for $p^{DC}(k_0,k_1,n)$,   it would be also useful to control that  the unconditional  probability of inconclusive and misleading evidence is fairly low.
%
%\citet{De:Santis:2004} argues  that in his experience often  $p^{DC}>\zeta$ was enough to  bound the previous two probabilities.

Recall that $p^{DC}(k_0,k_1,n)$ is a weighted mixture of two components.
%It is thus implicit in the
%adopted SSD criterion that the aim of the experiment is production of substantial correct evidence,
%without privileging  either  hypotheses.
Accordingly,   criterion \eqref{eq:optim:sample:size:general} is not suitable if the  aim is to
control one of the two probabilities of correct and decisive evidence rather than the average. This can be the case in clinical trials,  where  interest centers on one  hypothesis,  $H_i$ say.
In this case it seems  more appropriate to select the optimal sample size $n_i^*$ by controlling directly    $p_i^{DC}$.
%\purple forse aggiungere qualcosa sul fatto che l'impstazione di De Santis è generica, ossia non dipende dal BF.

\black
\citet{Schonbrodt:Wagen:2017} also rely on the BF to
plan a design  %which produces compelling evidence, that is detecting
to detect with high probability
an effect when it exists. In our setting this  corresponds to decisive and correct evidence in favor of the alternative hypothesis when the  null  represents absence of an effect.
%-------------------
Similarly to \citet{Weiss:1997}, they demonstrate the usefulness of plotting the distribution of the BF under the null, as well as under the alternative hypothesis. %; see their Figure 3.
%Computations are done via simulation
%  and using a point-mass design prior on a fixed value of the alternative for simplicity of the exposition, and a default proper density for the analysis prior. SSD can  then be achieved essentially by trial-and-error
Computations are performed based on simulations
in a fixed-$n$ design,  although  an open-ended sequential design as well as a sequential design with maximal $n$ are considered.

More recently \citet{Pan:Banerjee:2021}  attempt to provide a simulation-based framework
for Bayesian SSD
making explicit use of  design and analysis priors.
Working primarily in the setting of conjugate Bayesian linear regression models,
the required computational power for SSD is relatively modest. They also show that several frequentist results can be obtained as special cases of their general Bayesian approach.

%Recent work on replication studies, from a Bayesian perspective. Micheloud and Held (2022 Stat Sci); Held Micheloud and Pawel 2021 ArXiv;  then also  Pawel et al. Do not cover replication studies in general; only those dealing with SSD.

\black
\subsection{Contribution and structure of the paper}
\label{subsection:highlights}

In this paper we consider the issue of
causal discovery through interventions.
% Unlike current algorithmic approaches to active learning,  we provide methodology for    sample size determination for achieving compelling evidence, diversamente da quanto fatto finora dove ci si limita ad lgoritmi for DAG identiifcation.
Current algorithmic approaches to active learning do not determine the sample size of the interventional data needed to reach
a desired level of statistical accuracy. Using ideas from
Bayesian experimental design, we determine,
at each intervention,
the minimal
sample size
guaranteeing
a pre-experimental  overall probability of decisive and correct evidence which is sufficiently large.
\black
Specifically,
we frame the problem of edge orientation as a comparison between two competing causal DAGs, and adopt the Bayes factor as a measure of evidence.
%Our end-result is an algorithm that, for each given variable in an input sequence, identifies
%namely the minimum number of observations needed to  assure a good inferential performance of the causal learning process.

%We tackled this problem from a Bayesian experimental design perspective by adopting the Bayes Factor as a measure of evidence between competing causal structures (model hypothesis). For any given sequence of manipulated variables our method determines the corresponding optimal collection of sample sizes. The latter guarantees a pre-determined degree of assurance that the intervention experiment will produce decisive and correct evidence in favor of the true causal-model hypothesis, therefore being able to correctly recover the underlying causal DAG structure.

%\blue forse rimpolpare un po' quanto sopra scritto

\black

The rest of this paper is organized as follows. In Section
\ref{sec:BF:edge:orientation}
we discuss the problem of testing edge orientation between two Gaussian DAG models.
We then compute the corresponding Bayes factor
and derive its predictive distribution under each of the two hypotheses.
The previous result is adopted for sample size determination in the active learning procedure presented in Section
\ref{sec:Bayes:SSD:active:learning}.
The latter is illustrated through simulations and applied to a real dataset in Section
\ref{sec:illustrations}.
Finally in Section
\ref{sec:discussion}
we  analyze some critical points and  discuss new settings of application of the proposed methodology.
A few technical results relative to priors for DAG-model parameters and computations of Bayes factors are reported in the Appendix.
\black

\section{Bayes factor for edge orientation in Gaussian DAGs}
\label{sec:BF:edge:orientation}

\subsection{DAGs, Markov equivalence and interventions}
\label{sec:dags:markov:interventions}
\begin{figure}
	\begin{center}
		\begin{tikzpicture}
			\vspace{0.5cm}
			\begin{scope}[every node/.style={circle,thick,draw}]
				\node (1) at (0,0) {1};
				\node (2) at (1.8,0) {2};
				\node (3) at (0.9,-1.6) {3};
				\node (4) at (3.6,0) {4};
				\node (5) at (2.8,-1.6) {5};
			\end{scope}
			
			\begin{scope}[>={Stealth[black]},
				every node/.style={fill=white,circle},
				every edge/.style={draw=black, thick}]
				\path [-] (1) edge (2);
				\path [-] (2) edge (3);
				\path [-] (1) edge (3);
				\path [-] (4) edge (5);
			\end{scope}
			\begin{scope}[>={Stealth[black]},
				every node/.style={fill=white,circle},
				every edge/.style={draw=black, thick}]
				\path [->] (2) edge (4);
				\path [->] (2) edge (5);
			\end{scope}
		\end{tikzpicture}
	\end{center}
	\caption{A CPDAG with two chain components $\tau_1=\{1,2,3\}, \tau_2=\{4,5\}$. Edges between chain components are directed, while edges linking nodes belonging to the same chain component are undirected.}
	\label{fig:cpdag}
\end{figure}
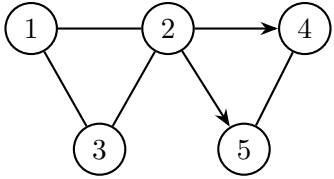

Let $\D=(V,E)$ be a Directed Acyclic Graph (DAG) whose vertices $V=\{1,\dots,q\}$ correspond to variables $Y_1,\dots,Y_q$ and $E\subseteq V \times V$ is the set of directed edges.
A DAG encodes a set of conditional independence relations between variables which can be read-off from the DAG, e.g. by using \textit{d-separation} \citep{Pear:2000}. We assume that an observational dataset $\bZ$ is available, where
\be
\bZ = \left(
\begin{array}{c}
	\bz_1^\top\\
	\bz_2^\top\\
	\vdots \\
	\bz_N^\top\\
\end{array}
\right),
\ee
with $\bz_i=(z_{i,1},\dots,z_{i,q})^\top$ for $i=1,\dots,N$.
In general, based on  observational data, $\D$ is identifiable only up to its Markov equivalence class $[\D]$, which collects all DAGs sharing the same conditional independencies.
Such DAGs are characterized  by having the same skeleton (the underlying undirected graph obtained by  disregarding  edge orientation) and \textit{v}-structures (sub-graphs of the form $u \rightarrow v \leftarrow z$ with $u$ and $z$ not connected)   \citep{Verma:Pearl:1990}.
\black
%\purple [dire "assuming faithfulness?" o che vale almeno nel gaussiano "classico"?] \black
Moreover, each equivalence class can be uniquely represented by a \textit{partially} directed acyclic graph named Essential Graph (EG) \citep{Ande:etal:1997} or Completed Partially Directed Acyclic Graph (CPDAG) \citep{Chic:2002}.
Let $\mathcal{E}(\D)\equiv\mathcal{E}=(V,E_{\E})$ be the CPDAG representing $[\D]$.
\citet{Ande:etal:1997} show that $\mathcal{E}$ is a \textit{chain graph} with \textit{decomposable} chain components.
We let $\mathcal{T}$ be the set of chain components of $\mathcal{E}$, with element $\tau\in\mathcal{T}$, and $\mathcal{E}_{\tau}=(\tau,E_{\tau})$ the \textit{sub-graph} of $\mathcal{E}$ induced by $\tau$, where $E_{\tau}=\{(u,v)\in E_{\E} \g u,v \in \tau\}$. \black
Importantly, $\mathcal{T}$ defines a partition of $V$, and   each chain component corresponds to an \textit{undirected} decomposable graph, while edges between nodes belonging to distinct chain components are directed; see also Figure \ref{fig:cpdag} for a simple example.
%%%%%%%%%

Under DAG $\D$, the joint density of $\bY=(Y_1,\dots,Y_q)$ factorizes as
\begin{equation}
	\label{eq:DAG:factorization}
	f(\by \g \D)=\prod_{j=1}^q f(y_j\g\by_{\pa_\D(j)}),
\end{equation}
where $\pa_\D(j)$ is the set of parents of node $j$ in $\D$ and $\by_A$ is the vector of variables representing nodes in $A \subseteq V$.
\black
Consider now an intervention on $Y_u$, $u \in V$, as obtained from a randomized experiment \citep{He:etal:2008} which replaces
$Y_u$ with a new r.v. $\widetilde{Y}_u$ having density $\widetilde{f}_u(\cdot)$.
We call $Y_u$  the manipulated variable (also named \textit{intervention target}) \citep{Haus:Buhl:2012},
and the \textit{do-operator} \citep{Pear:2000} is used to denote such an intervention.
The \textit{post-intervention} joint distribution of
$\bY$
is defined as
\begin{equation}
	\label{eq:int:distr}
	%\nonumber
	f(\by\g \textnormal{do}(Y_u=\widetilde{Y}_u), \D)=\tilde{f}_u(y_u)\prod_{j \ne u}f(y_j\g\by_{\pa_\D(j)}).
\end{equation}
Notice that the $f(y_j\g\by_{\pa_\D(j)})$'s
are the pre-intervention densities appearing in \eqref{eq:DAG:factorization}.
\black
As discussed in Section \ref{sec:intro:DAG:identification:interventions},
interventional data, namely those produced after an intervention on a variable among $Y_1,\dots, Y_q$, %as in \eqref{eq:int:distr},
can be used to identify the orientation of an undirected edge in $\E$.
%Let $e(u)$ be the edge orientations determined by manipulating node $u$.
Specifically, let $\E$ be a CPDAG and suppose that the undirected edge $u - v$ occurs in $\E$.
This implies that there are two DAGs, $\D_0$ and $\D_1$, in the Markov equivalence class of $\E$ which contain $u \leftarrow v$ and $u \rightarrow v$ respectively.
From \eqref{eq:int:distr} one can show that,  following an \emph{intervention} on $Y_u$,
\ben
\label{eq:H0:H1:dags}
Y_u \ind Y_v \,\, \textnormal{under} \,\, \D_0, \quad
Y_u \Neg{\ind} Y_v \,\, \textnormal{under} \,\, \D_1.
\een
The result follows using using \textit{d-separation}  because $u$ and $v$ are \emph{separated}  in the moral graph of the {ancestral} set of $\{  u,v\}$ \citep[Sect. 5.3]{Cowe:Dawi:Laur:Spie:1999};
see also \citet{He:etal:2008}.
%per capire meglio: Y_u in Y_v richiede che u e v siano separati nel grafo morale di An(u union v). Ma sotto (D_0,do u) u ha solo figli ed è separato da v e dai suoi genitori. Viceversa sotto (D_1, do u) u rimane genitore di u e quindi e' incluso negli ancestor di (u,v)

In principle, performing \textit{multiple} interventions followed by independence tests  in post-intervention distributions, one can  recover a  DAG structure by orienting all those edges that are undirected in $\E$.
\black
%\noindent Let now $[\D]$ be the Markov equivalence class of $\D$, $\G$ the representative EG. We denote with$[\D]_{e(i)}$ the subclass of $[\D]$ containing all DAGs $\D\in[\D]$ that share the same orientation $e(i)$. Then, for any $i - j$ in $\G$, with an intervention on node $i$, the Markov equivalence class of $\D$, $[\D]$ is reduced to the post intervention Markov equivalence class $[\D]_{e(i)}$.
%As shown in \citet{Haus:Buhl:2012}, each Markov equivalence class $[\D]_{e(i)}$ can be represented by a chain graph obtained as the union of all $\D\in[\D]_{e(i)}$ which still has the properties of EGs; see Theorem \ref{charact}. This is called \textit{Interventional Essential Graph} (I-EG).
%-------
The active learning approach of \citet{He:etal:2008} is based on a repeated  use of \eqref{eq:H0:H1:dags}. Specifically, it starts from an input Markov equivalence class (estimated from an observational dataset $\bZ$) and then selects interventions according to an optimal strategy which minimizes the number of manipulated
variables that are needed to achieve DAG identification.
\black

\subsection{Analysis prior and Bayes factor computation}
\label{sec:analysis:prior:BF}

In this section we first consider a Bayesian model for the observations conditionally on an input CPDAG $\E$. Next we derive
the Bayes Factor (BF) between two specific DAG models belonging to the equivalence class represented by $\E$. The resulting BF is used in the testing procedure \eqref{eq:H0:H1:dags} which underlies the approach to  \black sample size determination we describe in Section \ref{sec:Bayes:SSD:active:learning}.
%\purple [va detto qualcosa su compatibility, nel senso che si parte da un modello su un EG (o sua chain component) e si deriva compatibilmente il modello sul DAG (chain component orientata)? Ora vedere anche dettagli in Appendix] \black

Under a chain graph $\mathcal{E}$ the joint density of $\bY$ %$Y_1,\dots,Y_q$
factorizes \citep{Andersson:et:al:2001} as
\ben
\label{eq:fact:chain:graph}
f(\by\g\btheta_{\mathcal{E}})
=\prod_{\tau\in\mathcal{T}}
f_{\tau}(\by_{\tau}\g\by_{\pa_{\E}(\tau)},\btheta_{\tau}),
\een
%where $\pa_{\E}(\tau)$ is the set of parents of nodes in $\tau$ in $\E$ and $\bx_{A}$ denotes the sub-vector of $\bx$ with components indexed by $A\subseteq V$.
where  $\btheta_{\E}=\left\{\btheta_{\tau},\tau\in\mathcal{T}\right\}$
is a parameter indexing the graphical model $\E$. A specific feature of $\E$ is that \emph{all} nodes in ${\tau}$ share the same parents $\pa_{\E}(\tau)$
\citep[Thm 4.1 (iii)]{Ande:etal:1997}.
Since parameters $\btheta_{\tau}$'s are variation independent
\citep{Drton:Eichler:2006},
we will further assume that the prior on $\btheta_{\E}$ factorizes as
\ben
\label{eq:global:par:ind}
p(\btheta_{\E})=\prod_{\tau\in\mathcal{T}} p(\btheta_{\tau}),
\een
a condition which can be named \textit{global} parameter independence following \citet{Cast:Perl:2004}.

To recover a DAG structure from $\mathcal{E}$,
%\purple[in realtà il ns principale scopo è SSD, però per selezionare le target guardiamo alla DAG identification]\black
we need to determine the orientation of all the undirected edges in $\mathcal{E}$.
Since each undirected edge $u-v$ belongs to one chain component only, say $\tau$, we can restrict our attention to $\mathcal{E}_\tau$, the undirected decomposable graph of chain component $\tau$, and work separately on each chain component because of factorizations \eqref{eq:fact:chain:graph} and \eqref{eq:global:par:ind}. A further useful feature, highlighted in \citet[Thm 4]{He:etal:2008}, is the following:  if neither cycles nor     \textit{v}-structures are created during the process of edge orientation in a given chain component, then neither cycles nor \textit{v}-structures are introduced in the whole graph, too. Moreover, because a CPDAG is uniquely characterized by its skeleton  and \textit{v}-structures \citep{Ande:etal:1997}, %this feature guarantees that
any DAG obtained by orienting the original CPDAG $\E$ as described above in the previous paragraph still belongs to the equivalence class of $\E$.
%\grey "Given an essential graph G*, we need to orient all undirected edges in each chain component
%to discover the whole causal graph G. Below we show that the orientation can be done separately
%in every chain component. We also show that there are neither new v-structures nor cycles in the
%whole graph as long as there are neither v-structures nor cycles in any chain component. Thus in
%the orientation process, we only need to ensure neither v-structures nor cycles in any component,
%and we need not check new v-structures and cycles for the whole graph.".
Consider the orientation of  edge $u-v$ with $u, v \in \tau$. Write for simplicity   $\E_\tau \equiv \G$. From  \eqref{eq:H0:H1:dags} we deduce that independence holds if $u \leftarrow v$, while $u \rightarrow v$ otherwise.

\black
To determine edge orientation  we first write explicitly the general term $f_{\tau}(\cdot)$ in \eqref{eq:fact:chain:graph} using the standard factorization of the joint distribution  for decomposable graphical models \citep{Laur:1996}.
For better clarity, we use $X$ for the variables in  chain component $\tau$, and  denote $\left\{Y_j, j \in \tau\right\}$ with $\{X_1,\dots,X_T\}$, where $T=|\tau|$.
%A particular class of undirected graphs is represented by \textit{decomposable} UGs, also called \textit{chordal} or \textit{triangulated}; see for instance \citet{Laur:1996}. An undirected graph is decomposable if every cycle of length $l\ge 4$ has a \textit{chord}, that is two non-consecutive adjacent vertices. For a decomposable graph $\G$ on the set of vertices $V$, a complete subset that is maximal with respect to inclusion is called a \textit{clique}; see for instance graph $\G$ in Figure \ref{decomposable}.
Let also $\mathcal{C} = \{C_1,\dots, C_K\}$ be a \textit{perfect} sequence of cliques of the decomposable graph $\G$ \citep[p.~18]{Laur:1996}. Consider now, for $k=2,\dots, K$, the three types of sets
\be
H_k &=& C_1 \cup \dots \cup C_k, \\
S_k &=& C_k \cap H_{k-1}, \\
R_k &=& C_k \setminus H_{k-1},
\ee
which are called \textit{history}, \textit{separators} and \textit{residuals} respectively, and set $R_1=H_1=C_1, S_1=\emptyset$. Note that $C_1 \cup R_2 \cup \dots \cup R_K = V$ and also $R_k \cap R_{k'}=\emptyset$. It is then possible to number the vertices of a decomposable graph starting from those in $C_1$, then those in $R_1, R_2$ and so on. In this way we obtain a \textit{perfect numbering of vertices}, and a \textit{perfect directed version} $\G^{<}$ of $\G$,  by directing its edges from lower to higher numbered vertices.
Hence, we can write
\ben
\label{eq:fact:chain:residuals}
f(\bx\g\btheta_{\G^<}) &=&
\prod_{k=1}^{K}
f(\bx_{R_k}\g \bx_{S_k},\btheta_{R_k});
\een
%where $R_{k,j}$ is the \textit{j}-th element in $R_k$
see \citet[Eq.~35]{Dawid:Lauritzen:1993}.
The \textit{k}-th term in \eqref{eq:fact:chain:residuals} can be further written (omitting subscript $k$ to ease  notation) as
\ben
\label{eq:factorization:Rk}
f(\bx_R\g\bx_S,\btheta_R)
=
\prod_{l=1}^{|R|}
f(x_{R,l}\g x_{R,1},\dots,x_{R,l-1}, \bx_S, \theta_{R,l}), \nonumber \\
\een
\black
where $x_{R,l}$ is the \textit{l}-th term of $\bx_R$.
Importantly, the previous decomposition holds for any ordering $(x_{R,1},\dots,x_{R,|R|})$ of $\bx_R$.
Also, we can always choose clique $C_1$ to be that which contains edge $u-v$ \citep[Lemma 2.18]{Laur:1996}. Now consider two perfect directed versions of $\G$:
\begin{itemize}
	\item $\G_0^{<} \equiv \D_0$, containing $u \leftarrow v$,
	\item $\G_1^{<} \equiv \D_1$ , containing $u \rightarrow v$,
\end{itemize}
%\purple [uso sempre 0-1 (anziché 1 e 2) cercando di avere coerenza col BF nella successiva; 0 fa sempre riferimento al DAG che "diventa" di indipendenza] \black
such that $\D_0$ and $\D_1$ are  identical except for the edges $u \leftarrow v$ and $u \rightarrow v$.

Consider now the assignment of a prior distribution on the parameter indexing $\D_i$, $i=0,1$.
We follow the general procedure of \citet{Geig:Heck:2002} for eliciting parameter priors under any DAG-model starting from a \emph{unique} prior on the parameter of a \emph{complete} DAG, wherein all vertices are linked so that no conditional independencies are implied.
The central idea is  that parameters indexing the same conditional distributions be given  identical priors under \emph{any} DAG, which in turn are derived from the unique prior  under a complete DAG.
Actually, this method is an effective way to build  compatible priors \citep{Dawid:Lauritzen:2001:compatibleprior,cons:vero:2008}  across models.
An important consequence of compatibility  is that \emph{marginal} data distributions (marginal likelihoods) will involve the distributions of vertices and neighbor variables derived from a single prior, thus dramatically simplifying the elicitation procedure.
More details on prior assignments are provided in  Appendix \ref{appA}.

Using \eqref{eq:fact:chain:residuals} and \eqref{eq:factorization:Rk} together with global parameter independence of the parameters $\big\{\btheta_{R,l} \big \}_{l=1}^{|R|}$,
the marginal data distribution under the two DAG models following an intervention on $X_u$ is respectively
\ben
f(\bx\g \textnormal{do}(X_u=\widetilde{X}_u),\D_0)
=\tilde{f}_u({x}_u) \, m(x_v)
\cdot \prod_{k=2}^Km(\bx_{R_k}\g\bx_{S_k}),
\label{eq:marginal:data:intervention:D0}
\\
f(\bx\g \textnormal{do}(X_u=\widetilde{X}_u),\D_1)
=\tilde{f}_u({x}_u) \, m(x_v\g x_u)
\cdot\prod_{k=2}^Km(\bx_{R_k}\g\bx_{S_k})
\label{eq:marginal:data:intervention:D1}
\een
%%%%%%%%%
where $\widetilde{X}_u \sim \tilde{f}_u(\cdot)$.   Recall that the conditional distributions $m(\cdot\g\cdot)$, as well as the marginal one,  appearing in the right-hand sided of each equation are derived from the same prior.
%on the parameter for any complete model.
Hence terms in \eqref{eq:marginal:data:intervention:D0} and \eqref{eq:marginal:data:intervention:D1} involving the same arguments will be identical.
Let now
\ben
\begin{aligned}
	\label{eq:def:H0:H1}
	H_0 &:  \textnormal{the interventional distribution is  \eqref{eq:marginal:data:intervention:D0} }
	\\
	H_1 &:  \textnormal{the interventional distribution is  \eqref{eq:marginal:data:intervention:D1} }
\end{aligned}
\een
Based on a sample of size $n$
\be
\bX^n = \left(
\begin{array}{c}
	\bx_1^\top\\
	\bx_2^\top\\
	\vdots \\
	\bx_n^\top\\
\end{array}
\right),
\ee
the  Bayes Factor of $H_0$ vs $H_1$ reduces to
\ben
\label{eq:BF:01:indep:vs:dep}
\begin{aligned}
	\textnormal{BF}^{n}_{01}(\bX_u^n,\bX_v^n) &= \frac{m(\bX_{v}^n)}{m(\bX_{v}^n\g\bX_u^n)} \\
	&= \frac{m(\bX_{u}^n)m(\bX_{v}^n)}{m(\bX_{v}^n,\bX_u^n)},
\end{aligned}
\een
where $\bX_u^n$ is the sub-vector of $\bX^n$ corresponding to column $u$.
Equation \eqref{eq:BF:01:indep:vs:dep}  reveals that testing for edge orientation of $u-v$ is equivalent to testing independence under the joint marginal $m(x_u,x_v)$ between data $\bX_u^n$ and $\bX_v^n$ observed after an intervention on $X_u$, in accordance with \eqref{eq:H0:H1:dags}.
Specifically, (post-intervention) independence corresponds  to  the  edge  $ u \leftarrow v$;   conversely  dependence to $ u \rightarrow v$.
%%%%%%%%ABBIAMO rinosso frase successiva per evitare confusione%%%%%%%%%%%%%%%%%%%%
%%%%%%%%%%%%%%%%%%%%%%%%%%%%%%%%%%%%%%%%%%%%%%%%%%%%%%%%%%%%%%%%%%%%%%%%%%%%%%%%%%%
%%%%%%%%%%%%%%%%%%%%%%%%%%%%%%%%%%%%%%%%%%%%%%%%%%%%%%%%%%%%%%%%%%%%%%%%%%%%%%%%%%%
%It is  straightforward  to show that  following an intervention on $v$,  the newly obtained BF %, $\widetilde{\textnormal{BF}}_{01}$ say,
%would be equal to
%${m(\bX_{v}^n,\bX_u^n)}/[{m(\bX_{u}^n)m(\bX_{v}^n)}]$, which again is a test of independence
%under the joint marginal $m(x_u,x_v)$. However independence now corresponds to
%$ u \rightarrow v$, while  dependence is paired with $ u \leftarrow v$.
%Notice however that
%the data $(\bX_u^n,\bX_v^n)$ are generated following an intervention on $X_v$, while
%in
\eqref{eq:BF:01:indep:vs:dep} they follow an intervention on $X_u$.

\black

\subsection{Predictive distribution of the Bayes factor for Gaussian DAG models}
\label{sec:BF:predictive:Gaussian}

It is important to realize that,  from an experimental design perspective,  the BF in \eqref{eq:BF:01:indep:vs:dep}
is a function of  (interventional) observations  $\bX^n$ yet to be collected;  hence it is a random variable whose distribution can be derived from the  \textit{predictive}
distribution of $\bX^n$  conditional on the available past observational data $\bZ$.

For a given chain component $\tau$, we assume that for
$n$ observations $\bx_i=(x_{i,1},\dots,x_{i,T})^\top$, $i=1,\dots,n$,
\ben
\begin{aligned}
	\label{eq:gaussian:model:text}
	\bx_1,\dots,\bx_n \g \bOmega &\overset{\textnormal{iid}}{\sim}
	\N_T\left(\bzero, \bOmega^{-1}\right), \, \bOmega \in {\mathcal P_{\G}},
\end{aligned}
\een
where ${\mathcal P_{\G}}$ is the space of symmetric and positive definite matrices
Markov w.r.t. the decomposable graph $\G$.
%------------
%Si cita/rimanda a G\&H e Consonni La Rocca 2012 per la costruzione della prior partendo dal modello completo; quanto dobbiamo dettagliare?]
We show  in Appendix \ref{appA} that,  based on an objective prior approach,
\eqref{eq:BF:01:indep:vs:dep} takes the value
\ben
\label{eq:BF:uv}
\textnormal{BF}_{01}^{n}(\bX_u^n,\bX_v^n) =
g(n)\left[1-(r_{uv}^n)^2\right]^{\frac{n-1}{2}},
\een
where
\be
g(n)=\frac{n}{\sqrt{\pi}}\frac{\Gamma\left(\frac{n}{2}\right)}{\Gamma\left(\frac{n+1}{2}\right)}
\ee
and $r^n_{u,v}$ is the sample correlation coefficient between  $\bX^n_u$ and $\bX^n_v$,
\ben
\label{eq:sample:corr:coeff}
(r_{uv}^n)^2 =
\frac{\left[(\bX_u^n)^\top(\bX_v^n)\right]^2}
{(\bX_u^n)^\top(\bX_u^n) \cdot (\bX_v^n)^\top(\bX_v^n)},
\een
which can be written as
\be
(r_{uv}^n)^2 =
\frac{\left[\sum_{h=1}^n x_{h,u}x_{h,v}\right]^2}
{\sum_{h=1}^n x_{h,u}^2 \sum_{h=1}^n x_{h,v}^2},
\ee
where $x_{h,u}$ is the $h$-component of vector $\bX_u^n$.

To perform Bayesian SSD we need to compute the posterior predictive distribution of $\textnormal{BF}_{01}^n(\bX_u^n,\bX_v^n)$ under the two model hypothesis $H_0$, $H_1$.
To ease  notation we simply write $\textnormal{BF}_{01}^{n}$ instead of $\textnormal{BF}_{01}^{n}(\bX_u^n,\bX_v^n)$ for the remainder of this section.
Since the BF in \eqref{eq:BF:uv} depends on the data through the sample correlation coefficient $r_{u,v}^n$, we can derive first the posterior predictive distribution of $r_{u,v}^n$ and then obtain the corresponding distribution for the BF.

\black
\subsubsection{Posterior predictive under $H_0$.}

\black
Recall that under DAG $\D_0$ and an intervention on variable $X_u$ we have $X_u \ind X_v$, so that the post-intervention model distribution of $(X_u, X_v)$ can be written as
\be
\nonumber
p(x_u,x_v\g \bSigma_{\{u,v\},\{u,v\}},\textnormal{do}(X_u=\widetilde{X}_u), \D_0)
= \widetilde{f}_u(x_u)f(x_v\g\bSigma_{v,v}),
\ee
where
%$\gamma_u^2 = \var(X_u\g\widetilde{f})$,
$f(x_v\g\bSigma_{v,v})$ is $\N(0,\bSigma_{v,v})$.
%Since $X_v\g\bSigma_{\{u,v\},\{u,v\}},\textnormal{do}(X_u=\widetilde{x}), \D_0)$
%$X_v \g \bSigma_{v,v} \sim\N(0,\bSigma_{v,v})$.
Using Lemma 5.1.1 and Corollary 5.1.2 of \citet[p.~147]{Muirhead:1982} we obtain
\ben
\label{eq:dits:run:D0}
(r_{u,v}^n)^2\g\bSigma_{v,v},\textnormal{do}(X_u=\widetilde{X}_u), \D_0)
\sim \textnormal{Beta}\left(\frac{1}{2},\frac{n-1}{2}\right),
\een
so that $(r_{u,v}^n)^2$ is an ancillary statistic.
%\purple [occorre condizionare al do e D0? E' implicito che abbiano "reso" le due variabili indipendenti]
\black
As a consequence   \eqref{eq:dits:run:D0} coincides with  the posterior predictive distribution which we can simply write as $p\big((r_{u,v}^n)^2\g \textnormal{do}(X_u=\widetilde{X}_u), \D_0\big)$.
%becomes
%\ben
%\nonumber
%&&p\big((r_{u,v}^n)^2\g \textnormal{do}(X_u=\widetilde{x}), \D_0,\bZ\big)\\
%\nonumber
%&=&\int\int
%p\big((r_{u,v}^n)^2\g\bSigma_{v,v},\gamma_u^2,\textnormal{do}(X_u=\widetilde{x}), \D_0)\big)
%p(\sigma_v^2,\gamma_u^2\g\bZ)\, d\sigma_v^2\, d\gamma_u^2\\
%\nonumber
%&=&
%p\big((r_{u,v}^n)^2\g \textnormal{do}(X_u=\widetilde{x}), \D_0\big).
%\een
Hence, the posterior predictive of $\textnormal{BF}_{01}^n$ under
$H_0$ is analytically available and can be easily sampled from because
\ben
\label{eq:post:pred:BF:D0}
\begin{aligned}
	\textnormal{BF}_{01}^n &= g(n)\big(1 - (r_{u,v}^n)^2\big)^{\frac{n-1}{2}}
	\quad \textnormal{with}\quad\big(1 - (r_{u,v}^n)^2\big) \sim
	\textnormal{Beta}\left(\frac{n-1}{2},\frac{1}{2}\right).
\end{aligned}
\een
% is the variance of $X_u$ in the interventional density $\widetilde{f}$

\subsubsection{Posterior predictive under $H_1$.}
\label{subsec:posterior predcitive}

\black
Under DAG $\D_1$ and an intervention on variable $X_u$ the post-intervention model distribution of $(X_u,X_v)$ is
\be
\nonumber
p(x_u,x_v\g \bSigma_{\{u,v\},\{u,v\}},\textnormal{do}(X_u=\widetilde{X}_u), \D_1)
= \widetilde{f}_u(x_u)
f(x_v\g x_u, \bL_{u,v},\bD_{v,v}),
\ee
where
\ben
\label{eq:Luv:Dvv}
\quad\quad\bL_{u,v}=-\left(\bSigma_{\{u,v\},\{u,v\}}\right)^{-1}\bSigma_{u,v},
\quad
\bD_{v,v} = \bSigma_{v\g u}.
\een
\black
Letting  $\bZ$ be the $(N,T)$ matrix of available \emph{observational} data,
we choose as \emph{design} prior for $(\bL_{u,v},\bD_{v,v})$
the posterior $p(\bL_{u,v},\bD_{v,v}\g\bZ,\D_1)$ which can be derived from the posterior of
$\bOmega=\bSigma^{-1}$,    as we show in Appendix \ref{appA}.

Returning to the posterior distribution of the unconstrained matrix $\bOmega$,
%----------
%\purple
%obtained following the same elicitation procedure adopted in
%\eqref{eq:gaussian:model} and based on the method of \citet{Geig:Heck:2002}
consider the model and objective prior
\ben
\begin{aligned}
	\bz_1,\dots,\bz_N \g \bOmega &\overset{\textnormal{iid}}{\sim}
	\N_T\left(\bzero, \bOmega^{-1}\right)\\
	\nonumber
	\bOmega&\sim p(\bOmega) \propto |\bOmega|^{\frac{a_{\Omega}-T-1}{2}}.
\end{aligned}
\een
We obtain
\ben
\bOmega\g\bZ&\sim\W_T(a_{\Omega}+N,\bS),
\een
where $\bS=\bZ^\top\bZ$,  and this  acts as the generating {design} prior for the parameters in \eqref{eq:Luv:Dvv}.
Now
\ben
\nonumber
\left(\bSigma_{\{u,v\},\{u,v\}}\right)^{-1}
= \bOmega_{\{u,v\},\{u,v\}\g\tau\setminus\{u,v\}}
:= \boldsymbol{Q}_{u,v}
\een
so that
\be
\boldsymbol{Q}_{\{u,v\}}
\sim\W_2\left(a_{\Omega}+N-(T-2),\bS_{\{u,v\},\{u,v\}}\right),
\ee
using distributional properties of the Wishart distribution \citep[Thm 5.1.4]{Press:1982}.
Hence, the posterior predictive of $\textnormal{BF}_{01}^n$ under $H_1$ can be approximated by Monte Carlo simulation following Algorithm
\ref{alg:post:pred:BF:D1:alg}.

\black

\black
\begin{algorithm}{
		\SetAlgoLined
		\vspace{0.1cm}
		\KwInput{Observational $(N,T)$ data matrix $\bZ$;
			interventional density $\widetilde{f}_u(\cdot)$;
			prior hyperparameter $a_{\Omega}$;
			required sample size $n$; number of Monte Carlo draws $S$}
		\KwOutput{A sample of size $S$ from the posterior predictive distribution of $\textnormal{BF}_{01}^n$ under $H_1$}
		Compute $\bS=\bZ^\top\bZ$ \;
		\For{$s=1,\dots,S$}{
			Draw $\boldsymbol{Q}^{(s)}_{u,v}
			\sim\W_2\left(a_{\Omega}+N-(T-2),\bS_{\{u,v\},\{u,v\}}\right)$ \;
			Compute $\bL_{u,v}^{(s)}, \bD_{v,v}^{(s)}$ \;
			\For{$h=1,\dots,n$}{
				sample $x_u^{(h)(s)}\sim \widetilde{f}_u(\cdot)$ \;
				sample $x_v^{(h)(s)}\sim \N\left(\cdot\g-\bL_{u,v}^{(s)}x_u^{(h)(s)},\bD_{v,v}^{(s)}\right)$ \;
			}
			and obtain $\bX_u^{n(s)}=\big(x_u^{(1)(s)},\dots, x_u^{(n)(s)}\big)^\top$ \\
			and $\bX_v^{n(s)}=\big(x_v^{(1)(s)},\dots, x_v^{(n)(s)}\big)^\top$ \; \black
			Compute $\big(r_{uv}^n\big)^{2(s)}$ using $\bX_u^{n(s)},\bX_{v}^{n(s)}$ as in \eqref{eq:sample:corr:coeff} \;
			Compute $\textnormal{BF}_{01}^{n(s)}= g(n)\left[1-\big(r_{uv}^n\big)^{2(s)}\right]^{\frac{n-1}{2}}$
			as in \eqref{eq:BF:uv}
		}
		Return $\left\{\textnormal{BF}_{01}^{n(1)},\dots,\textnormal{BF}_{01}^{n(S)}\right\}$
	}
	\caption{Approximate posterior predictive of $\textnormal{BF}_{01}^n$ under $H_1$}
	\label{alg:post:pred:BF:D1:alg}
\end{algorithm}

%%%%%%%%%%%%%%%%%%%%%%%%%%%%%%%%%%%%%%%%%%%%%%%%%%%%%%%%%%%%%%%%

\section{Bayesian sample size determination for active learning}
\label{sec:Bayes:SSD:active:learning}

Let $\mathcal{E}$ be a CPDAG with set of chain components $\mathcal{T}$.
As in Subsection \ref{sec:analysis:prior:BF}, in the following we restrict our attention to a given chain component $\tau \in \mathcal{T}$ and let $\mathcal{E}_{\tau}\equiv \mathcal{G}$ be the corresponding (decomposable undirected) sub-graph.

\subsection{Single-edge orientation}

Consider an undirected edge $u-v$  in $\G$, whose orientation has to be determined.
We argued in Section \ref{sec:analysis:prior:BF} that this can be done by testing $H_0$ \emph{vs} $H_1$ defined in \eqref{eq:def:H0:H1} leading to the BF in \eqref{eq:BF:01:indep:vs:dep} which we write as $\textnormal{BF}_{01}^{n}$ for short.
%
%For given prior probabilities $p(H_0)$ and $p(H_1) = 1-p(H_0)$, the posterior of $H_0$ can be written as
%\ben
%\label{eq:post:prob:H0}
%\begin{aligned}
%	\Pr(H_0\g \bX_{uv}^n)
%	&=
%	\frac{p(H_0)m(\bX_{uv}^n\g H_0)}{\sum_{j \in \{0,1\}}p(H_j)m(\bX_{uv}^n\g H_j)} \\
%	&=
%	\frac{p(H_0)\BF_{01}^n}{p(H_1) + p(H_0)\BF_{01}^n}
%\end{aligned}
%\een
%and similarly for $H_1$,
%\ben
%\label{eq:post:prob:H1}
%\begin{aligned}
%	\Pr(H_1\g \bX_{uv}^n)
%	&=
%	\frac{p(H_1)\BF_{10}^n}{p(H_0) + p(H_1)\BF_{10}^n},
%\end{aligned}
%\een
%where $\BF_{10}^n = \left\{\BF_{01}^n\right\}^{-1}$.
%
%\brown la rimanente parte della subsection è già nell'introduzione; occorre decidere cosa fare
%
%\black
%Following \citet{De:Santis:2004} we consider \textit{decisive and correct} evidence (DCE) in favor of $H_j$ as the event $\Pr(H_j\g \bX_{uv}^n)\ge \gamma_j$, where $\gamma_0, \gamma_1 \in(0,1)$ are fixed thresholds, that is   DCE holds when the data provide substantial evidence in favor of the correct hypothesis. Because of \eqref{eq:post:prob:H0} and \eqref{eq:post:prob:H1} and letting $\omega = p(H_0)/p(H_1)$, it is easy to show that
%\ben
%\nonumber
%\Pr(H_0\g \bX_{uv}^n)\ge \gamma_0 &\iff& \BF_{01}^n \ge \omega\frac{\gamma_0}{1-\gamma_0} := k_0, \\
%\nonumber
%\Pr(H_1\g \bX_{uv}^n)\ge \gamma_1 &\iff& \BF_{01}^n \le \frac{1}{\omega}\frac{1-\gamma_1}{\gamma_1} := \frac{1}{k_1}.
%\een
%%The previous expressions relate the occurrence of DCEs to the distribution of BF.

Based on the analysis presented in Subsection \ref{subsec:Bayesian experimental design}
we can define
the conditional probabilities of Decisive and Correct Evidence (DCE) as
\ben
\begin{aligned}
	\label{eq:prob:DCE:H0:H1}
	p^{DC}_{0}(k_0,n) &= \Pr\left\{\BF_{01}^{n} \ge k_0 \g H_0\right\}, \\
	p^{DC}_{1}(k_1,n)  &= \Pr\left\{\BF_{01}^{n} \le 1/k_1 \g H_1\right\}.
\end{aligned}
\een
%where we also emphasize the dependence on the data $\bX_{uv}^n$.
Finally, the overall probability  is
\black
\ben
\label{eq:prob:DCE}
%p^{DC}(\bX_{uv}^n,k_0,k_1) &=& p(H_0)p^{DC}_0(\bX_{uv}^n,k_0) \\
%&+& p(H_1)p^{DC}_1(\bX_{uv}^n,k_1)
p^{DC}_{uv}(k_0,k_1,n) &=& \sum_{j\in\{0,1\}}p(H_j)p^{DC}_{j}(k_j,n)
\een
and the optimal sample size  to reach DCE at level $\zeta \in (0,1)$ is
\ben
\label{eq:optim:sample:size}
n^*_{uv} = \min\left\{n\in \mathbb{N} : p^{DC}_{uv}(k_0,k_1,n) \ge \zeta\right\}.
\een
%{eq:optim:sample:size:general}

\black

\subsection{Multiple-edge orientation and sequences of manipulated variables}
\label{sec:optimal:sequence}
Consider now the decomposable UG $\G$ corresponding to one chain component of CPDAG $\mathcal{E}$.
Interventions on variables in $\mathcal{E}$ can be used to identify the orientation of undirected edges within each chain component, as they \textit{break} the equivalence class represented by $\mathcal{E}$ into a collection of smaller (interventional) equivalence classes; see also Section \ref{sec:dags:markov:interventions}. Assuming faithfulness \citep{Spir:Glym:Sche:2000}, by manipulating a sufficiently large number of nodes, we can in principle  identify a DAG structure through  independence tests between each intervened variable and its neighbors.
\citet{He:etal:2008}
%propose an active learning approach for discovering causal structures of DAG models.
propose an optimal design strategy
%identify the true DAG generating model via interventions on variables. Given a collection of
%observational data, theyfirst estimate a Markov equivalence class and then orient
%the undirected edges in the representative EG by manipulating variables through
%external interventions. The objective is then to...
which minimizes the number of manipulated
variables that are needed to guarantee that an equivalence class is progressively partitioned into smaller classes, eventually  comprising a single DAG.%, which in turn implies  DAG identifiability.
%Let S = fV1; : : : ; Vkg be a sequence of manipulated nodes, [D] a Markov equivalence
%class. We say that S is sucient if we can identify one DAG from all the possible
%DAGs in [D] after the variables in S are manipulated. Furthermore, we say that
%S is minimal if any subset of S is not sufficient.
Specifically, a sequence of manipulated variables
$S = (u_1,\dots,u_K)$ is \textit{sufficient} for $\G$
if one can identify
a single DAG from all possible DAGs in $\mathcal{G}$ after variables in $\mathcal{S}$ are manipulated.
The optimal sequence of manipulated variables is then defined as follows.
\begin{definition}[\citet{He:etal:2008}]
	\label{def:optimal:he:geng}
	Let $S = (u_1,\dots,$ $u_K)$ be a sequence of manipulated variables.
	Then $S$ is optimal if $|S|=\min\{|S_l|: S_l \in \mathcal{S}\}$,
	where $\mathcal{S}$ is the set of all sufficient sequences.
\end{definition}
The resulting optimal sequence is not unique in general, as the  following example shows.

\vspace{0.5cm}

\textit{Example}. Consider a graph $\mathcal{G}: u - v$, representing a chain component. Both $S_1 = \{u\}$ and $S_2=\{v\}$ are sufficient sets of manipulated variables, because they allow to distinguish between $u \rightarrow v$ and $u \leftarrow v$. Since there are no other sufficient sets of smaller size, both $S_1$ and $S_2$ are also optimal according to Definition \ref{def:optimal:he:geng}.

\vspace{0.5cm}

For a given chain-component graph $\G$, consider  an optimal sequence $S=(u_1,\dots,u_K)$ as in Definition \ref{def:optimal:he:geng}.
Notice that each node $u \in S$ is typically linked to a number of nodes $v$ in the chain component, namely its \textit{neighbors}
of $u$ in $\G$, $\textnormal{ne}_\G(u)$.
Consider a node $v \in \textnormal{ne}_\G(u)$;
from \eqref{eq:prob:DCE} we need to assign $P(H_0)$ and $p(H_1)=1-P(H_0)$.
Recall that  $H_0$ corresponds to $u \leftarrow v$,  while the direction is reversed under $H_1$.
A way to proceed is to consider all DAGs which are perfect directed versions of $\G$ whose set is denoted by $[\G]$. Since observational data cannot distinguish among them, it is natural to regard them as equally likely.
Accordingly we set
\ben
\label{eq:prob:H0}
p(H_0) \propto \sum_{\D\in [\G]}\mathbbm{1}_{u \leftarrow v}\{\D\}, \quad
%\nonumber
%p(H_1) \propto \sum_{\D\in [\G]}\mathbbm{1}_{u \rightarrow v}\{\D\},
\een
where $\mathbbm{1}_{u \leftarrow v}\{\D\}=1$ iff $\D$ has $u \leftarrow v$, so that $p(H_0)$
is proportional to the number of DAGs in $[\G]$ containing $u \leftarrow v$.
%; similarly,
%$p(H_1)$ will be proportional to the number of DAGs in $[\G]$ containing $u \rightarrow v$.

From \eqref{eq:optim:sample:size} we can determine the optimal sample size $n^*_{uv}$
for each $v \in \textnormal{ne}_{\G}(u)$.
Furthermore, the optimal sample size for an intervention on $u$ becomes
\be
n^*_{u} = \max\left\{n^*_{uv},v \in \textnormal{ne}_{\G}(u)\right\}.
\ee

%\subsection{Algorithm}

For a given  sequence of manipulated variables,
our strategy for sample size determination is summarized in Algorithm \ref{alg:optimal:sequence:sample:size}.

\begin{algorithm}{
		\SetAlgoLined
		\vspace{0.1cm}
		\KwInput{A sequence of manipulated variables $S=(u_1,\dots,u_K)$,
			a threshold for probability of DCE $\zeta\in(0,1)$}
		\KwOutput{A collection of optimal sample sizes $\boldsymbol{n}^*$}
		\For{$u \in S$}{
			Construct the set of neighbors of $u$ in $\G$, $\textnormal{ne}_{\G}(u)$\;
			\For{$v \in \textnormal{ne}_{\G}(u)$}{
				Find $n^*_{uv} = \min\left\{n\in \mathbb{N} : p^{DC}_{uv}(k_0,k_1,n) \ge \zeta\right\}$\
			}
			Compute $n^*_{u} = \max\left\{n^*_{uv},v \in \textnormal{ne}_{\G}(u)\right\}$
		}
		Return $\boldsymbol{n}^*=(n_{u_1}^*,\dots,n_{u_K}^*)$
	}
	\caption{Optimal sequence of sample sizes}
	\label{alg:optimal:sequence:sample:size}
\end{algorithm}

Recall from Definition \ref{def:optimal:he:geng} that an optimal sequence of manipulated variables need not be unique, and define $\{S_1,\dots,S_L\}$ to be the \textit{collection} of all such sequences.
By applying Algorithm
\ref{alg:optimal:sequence:sample:size}
to a given sequence $S_l$, we obtain the corresponding vector of optimal sample sizes
$\boldsymbol{n}^{*(l)}=\big(n_{u_1}^{*(l)},\dots, n_{u_{K_l}}^{*(l)}\big)$ for which we can compute
the total sample size
\be
{N}^{*(l)} = \sum_{k=1}^{K_l}n_{u_{k}}^{*(l)}.
\ee
%Let $\{S_1,\dots,S_L\}$ be a collection of optimal sequences of manipulated variables as in Definition \ref{def:optimal:he:geng}.
Hence, the Best size Optimal Sequence of manipulated variables (BOS) is naturally defined as
the sequence $S^*$ having the smallest total sample size $N^*={\min}\left\{ N^{*(l)},l=1,\dots,L\right\}$.

%
%\begin{definition}[BOS]
%	\label{def:optimal:caste:cons}
%	Let $\{S_1,\dots,S_L\}$ be a collection of optimal sequences of manipulated variables as in Definition \ref{def:optimal:he:geng}.
%	Then the BOS  is
%	\ben
%	\nonumber
%S^*=\underset{\{S_l\}_{l=1}^L}{\arg \min}\left\{\widetilde N^{(l)},l=1,\dots,L\right\},	
%\een
%with corresponding sample size
%\ben
%N^*=\underset{\{N^{(l)}\}_{l=1}^L}{\arg \min}\left\{\widetilde N^{(l)},l=1,\dots,L\right\}.
%	\nonumber
%\een	
%	\end{definition}

\section{Illustration and real data analysis}
\label{sec:illustrations}

In this section we illustrate the proposed method on a simple example with chain components having two nodes  and apply it to  a high-dimensional  dataset
about riboflavin production by \textit{Bacillus subtilis}.

\black
\subsection{Two-node chain component}

Consider the  chain component $u - v$. The objective is to determine the optimal sample size for an intervention on $u$.
An observational dataset $\bZ$ is first generated as follows.
Assuming the true DAG generating model is $u \rightarrow v$, we consider the system of linear equations
\begin{equation}
\nonumber
\left\{\begin{array}{ll}
	X_u = \varepsilon_u &\quad \varepsilon_u \sim\N(0,1) \\
	X_v = 0.5 X_u + \varepsilon_u &\quad \varepsilon_v \sim\N(0,1)
\end{array}
\right.\,
\end{equation}
with $\varepsilon_u \ind \varepsilon_v$,  and then generate $N=50$ i.i.d. observations collected in the data matrix $\bZ$.

\begin{landscape}

\begin{figure*}%[b]
	\centering
\begin{tabular}{cc}
	\includegraphics[scale=0.51]{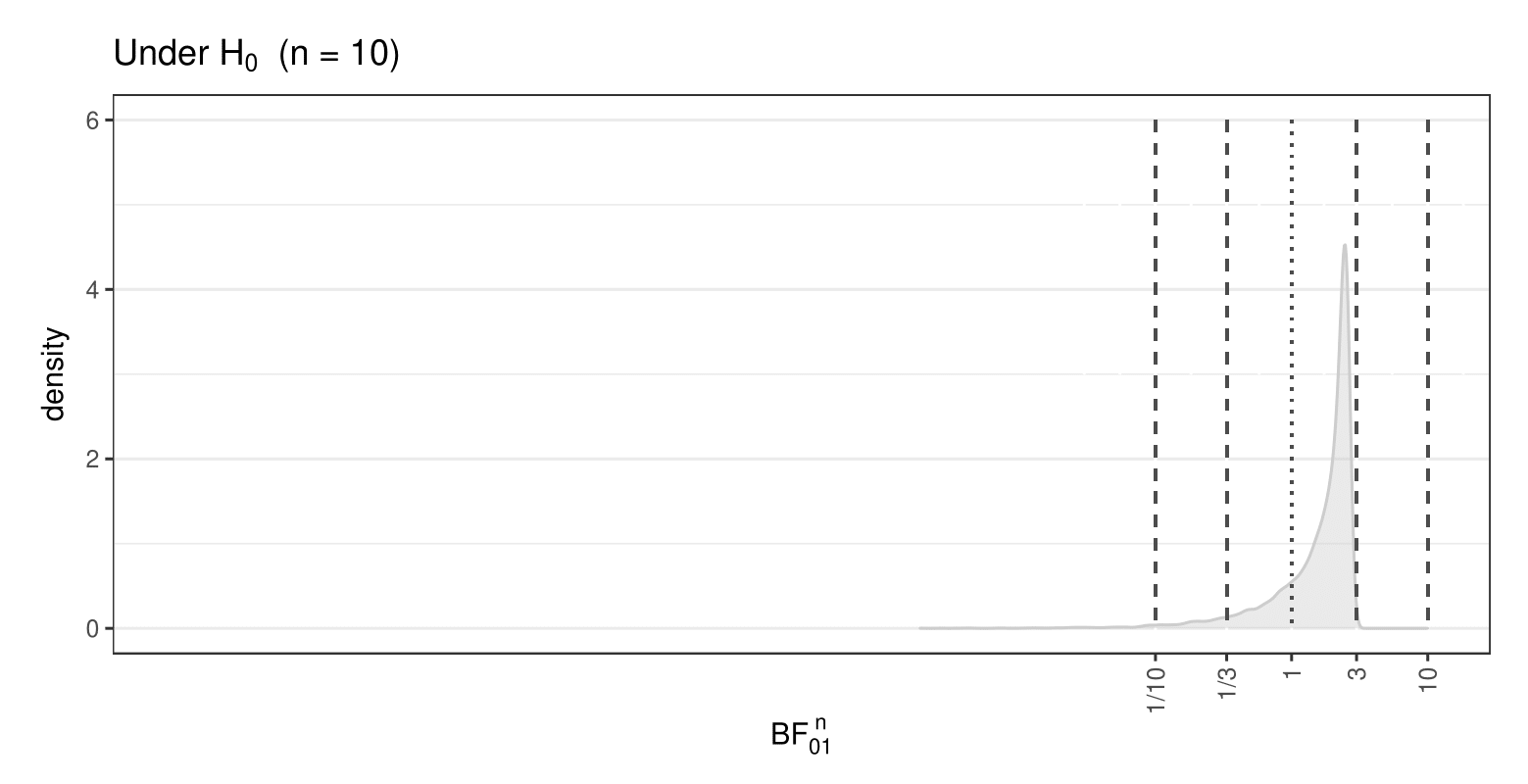} & \includegraphics[scale=0.51]{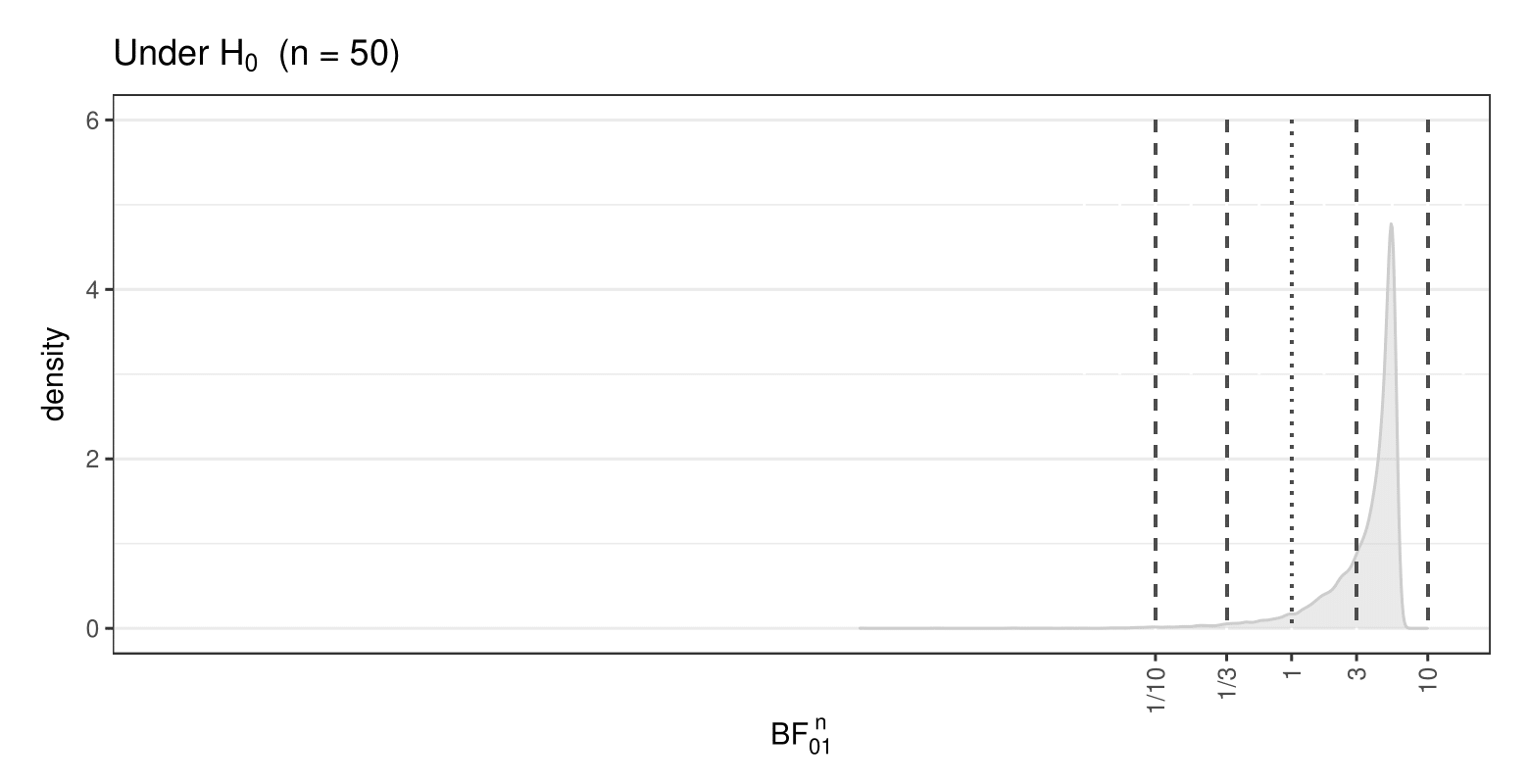} \\
	\includegraphics[scale=0.51]{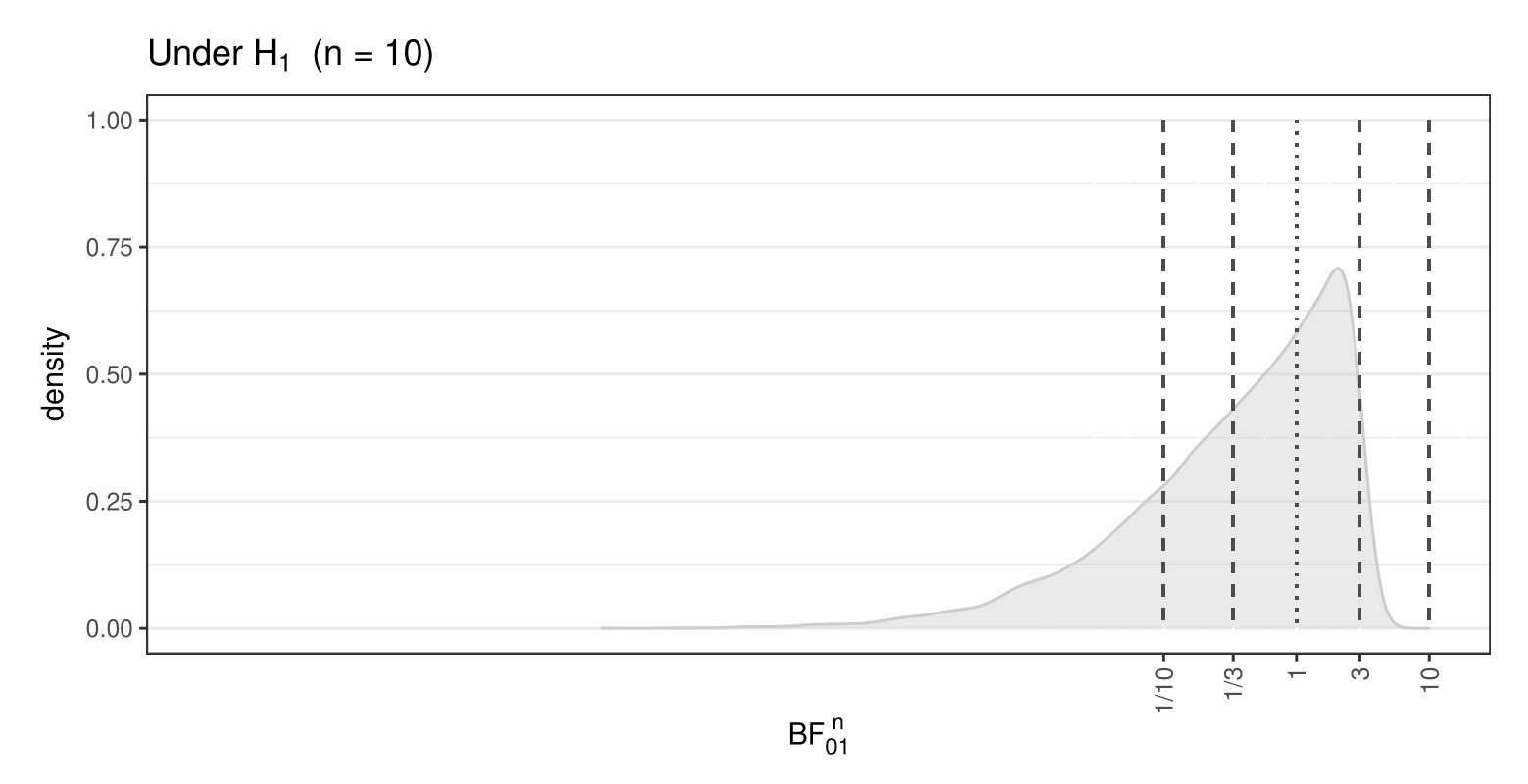} &
	\includegraphics[scale=0.51]{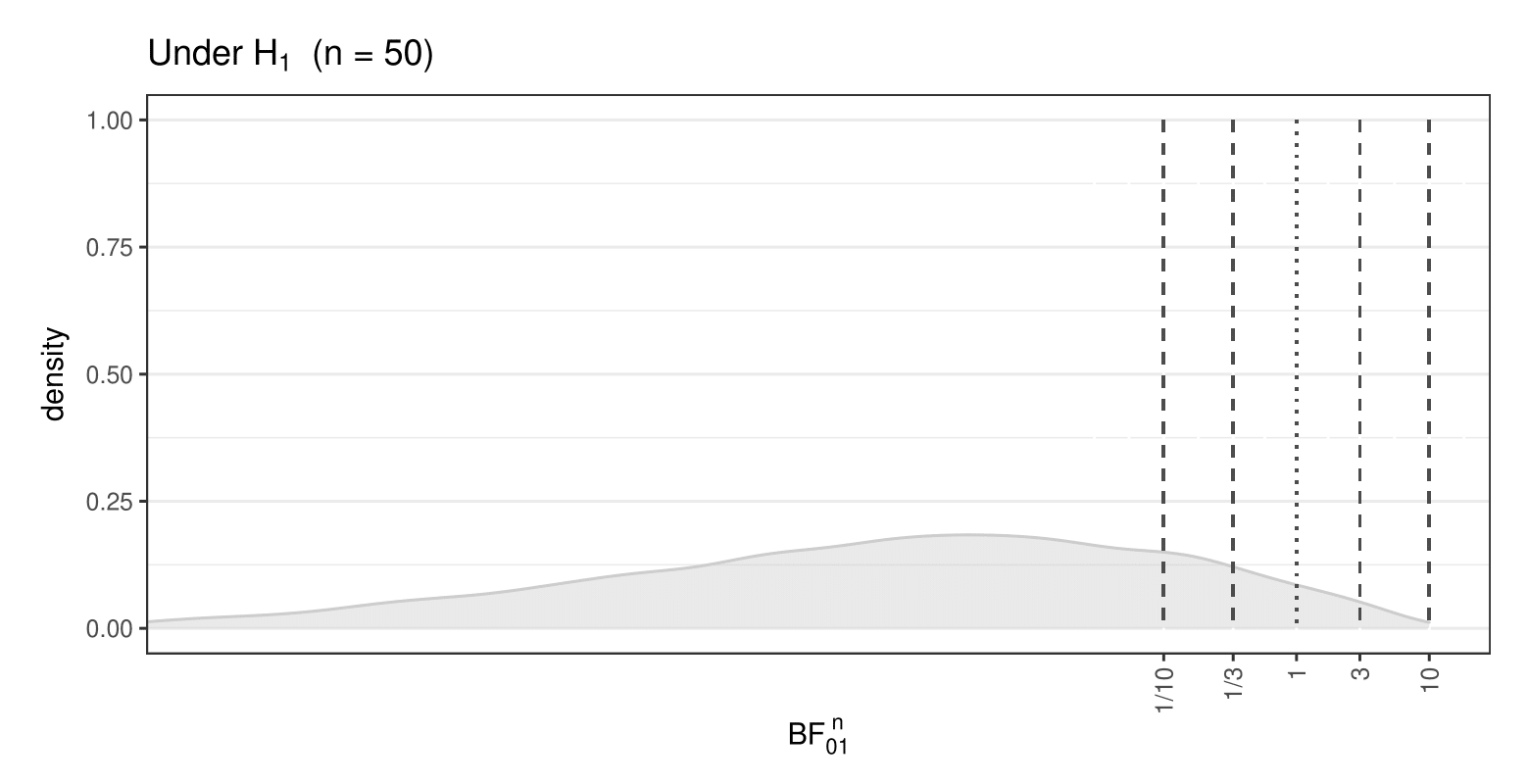}
\end{tabular}
\caption{Two-node chain component: Predictive distribution of
	%$\textnormal{BF}_{01}$
	$\textnormal{log$_{10}$ BF}_{01}^n$
	under $H_0$ and $H_1$ for sample sizes $n=10$ and $n=50$.
	Vertical lines represent BF thresholds $1/10,1/3,1,3,10$.}
\label{fig:sim:predictive:BF}
\end{figure*}

\end{landscape}

We first focus on the predictive distribution of $\textnormal{BF}_{01}^n$, the Bayes Factor
defined in \eqref{eq:BF:01:indep:vs:dep}.
Results are summarized in Figure \ref{fig:sim:predictive:BF}
which reports the (approximate) predictive distribution of   $\textnormal{log$_{10}$ BF}_{01}^n$ \black under each of the two hypotheses, for values of $n\in\{10,50\}$.
%------------
To ease legibility,  values on the horizontal axis are expressed as $\textnormal{ BF}_{01}^n$,
and thresholds corresponding to values in $\{1/10,1/3,1,3,10\}$ are reported as vertical lines.
\black
From this output, we can compute the probabilities  that the BF   favors  the true hypothesis for each of the degree  evidence categories in Table \ref{tab:BF:categ}.
Specifically, we focus on ``moderate evidence'' for $H_0$ and $H_1$, which corresponds to
$3<\textnormal{BF}_{01}^n<10$ and $1/10<\textnormal{BF}_{01}^n<1/3$, respectively.
We also consider   ``strong-to-extreme evidence''
for $H_0$ and $H_1$,
corresponding to
$\textnormal{BF}_{01}^n>10$  and $\textnormal{BF}_{01}^n<1/10$, respectively.
Results, for different sample sizes $n \in \{10,50,100\}$ are summarized in Table \ref{tab:sim:BF:probs}.
For $n=10$ the two probabilities are both zero under $H_0$, which is coherent with Figure \ref{fig:sim:predictive:BF} where the BF distribution does not exceed the threshold $3$.
By increasing the sample size to $n=50$ and $n=100$, the probability of moderate evidence increases up to about $84\%$, while the probability of strong-to-extreme evidence is only around 2\% for $n=100$.
Conversely, when the true hypothesis is $H_1$, we have strong evidence with a probability higher than $80\%$ even for a moderate sample size, $n=50$. The latter probability grows up to $96\%$ when the sample size increases to $n=100$.
We thus see an imbalance between the learning rate between $H_0$
and $H_1$, a phenomenon which is not new but still worth of consideration; see for instance \citet{Johnson:Rossell:2010}.

\begin{table}
\centering
\caption{Two-node chain component.  Predictive probabilities of  $BF_{01}^n$ resulting in  ``moderate'' or ``strong-to-extreme''   evidence in favor of the true hypothesis $H_0$ and $H_1$, for sample sizes $n \in \{10,50,100\}$.}
\label{tab:sim:BF:probs}
\begin{tabular}{cccccc}
	\hline
	True && $n$ && Moderate & Strong-to-extreme \\
	\hline
	&& $10$ && 0.00\% & 0.00\% \\
	$H_0$ && $50$ && 73.64\% & 0.00\% \\
	&& $100$ && 84.39\% & 2.10\% \\
	\hline
	&& $10$ && 18.5\% & 0.23\% \\
	$H_1$ && $50$ && 7.70\% & 82.9\% \\
	&& $100$ && 0.16\% & 96.2\% \\
	\hline
\end{tabular}
\end{table}

\begin{figure}%[b]
\centering
\includegraphics[scale=0.55]{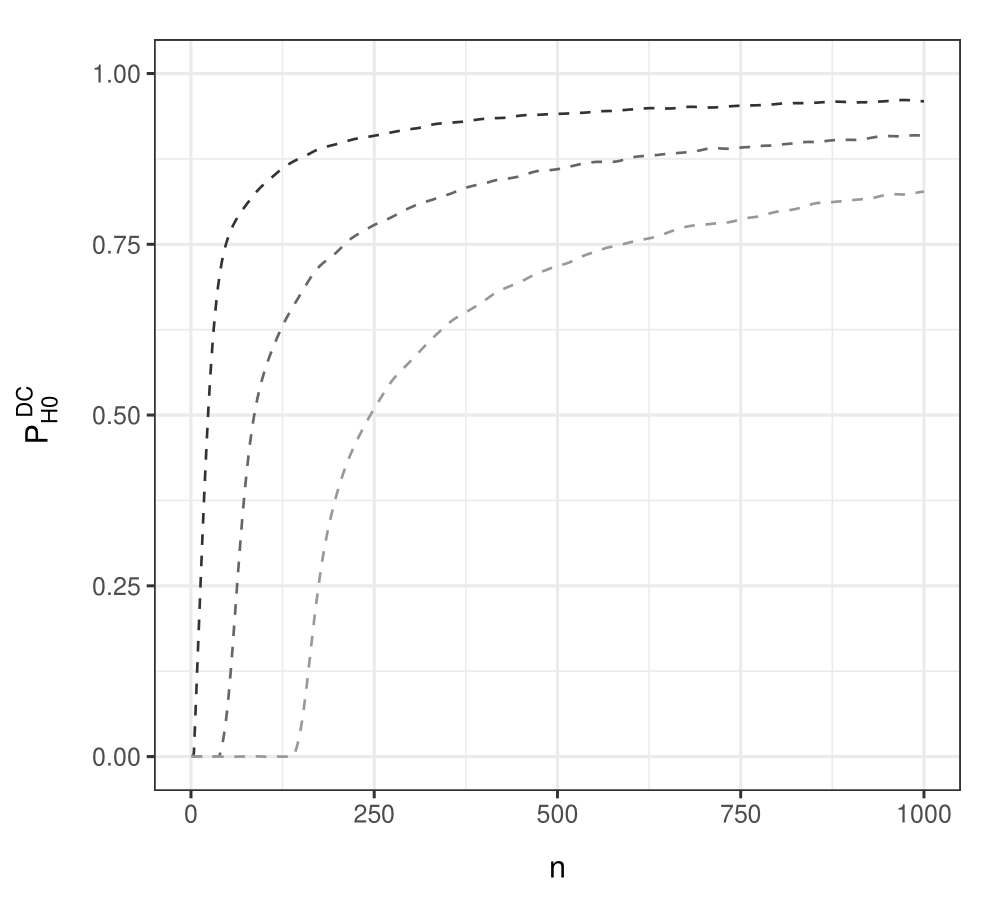} \\
\includegraphics[scale=0.55]{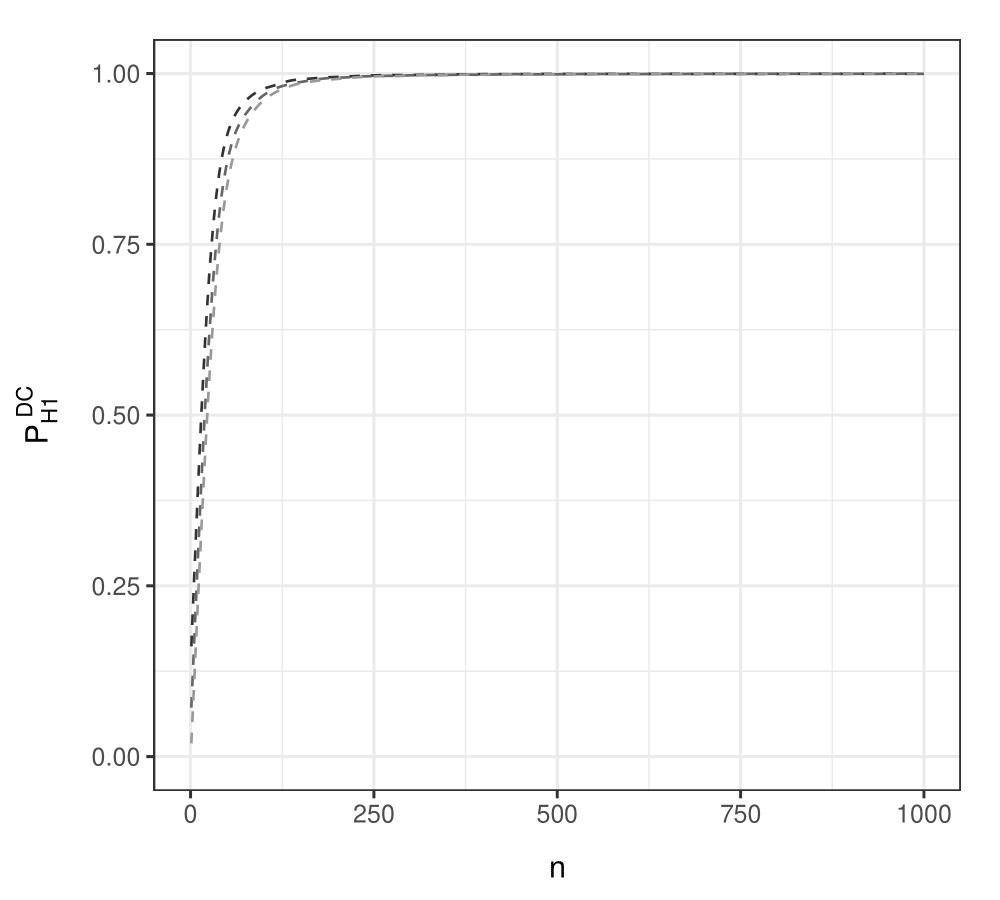} \\
\includegraphics[scale=0.55]{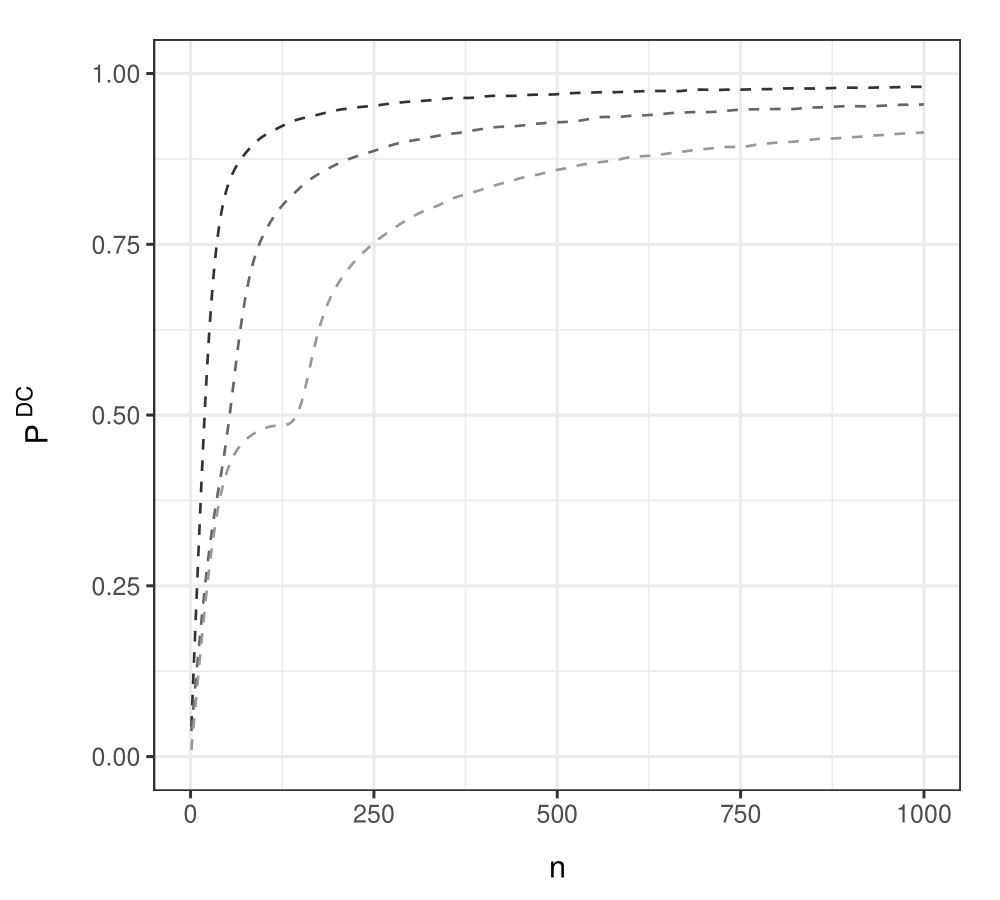}
\caption{Two-node chain component: Probability of Decisive and Correct (DC) evidence in favour of $H_0$, $H_1$ and overall probability (from top to bottom plots) as a function of the sample size $n$, for different values of $k\in\{3,6,10\}$ (from dark to light grey) and $k_0=k_1=k$.}
\label{fig:sim:prob:DCE}
\end{figure}

Consider now the probabilities of DCE as in Equations \eqref{eq:prob:DCE:H0:H1} and \eqref{eq:prob:DCE}.
We compute $p^{DC}_0(k_0,n)$ and $p^{DC}_1(k_1,n)$ for $k_0=k_1=k$ by varying $k\in\{3,6,10\}$, and for a grid of sample sizes $n\in\{1,2,\dots,1000\}$.
The behavior of the two probabilities as a function of $n$ is summarized in the first two plots of Figure \ref{fig:sim:prob:DCE} where each curve refers to one value of $k$ (from dark to light grey for increasing levels of the threshold).
Consider for instance $p^{DC}_0(k_0,n)$: this probability exceeds 80\% when  $k=3$ for a sample size  $n=100$, consistently  with the results of Table \ref{tab:sim:BF:probs}.
When instead $k=6$, the same sample size only guarantees that $p^{DC}_0(k_0,n)$ is  approximately equal to $65\%$; moreover, to reach a level of   $80\%$ the sample size must increase to about $n=400$.
Notice that $p^{DC}_0(k_0=10,n)$ is  zero for $n$ up to 150; this explains  the elbow  in the bottom panel of Figure \ref{fig:sim:prob:DCE}.

A similar behavior is observed for $p^{DC}_1(k_1,n)$, where however the distance between the three curves is much smaller, especially for moderate-to-large values of $n$.
This is coherent with the results in Figure \ref{fig:sim:predictive:BF} which suggest that the area to the left of 1/10 of the BF is already appreciable for small values of  $n$ such as 10.  In addition
the area to the right of $k=3$ or $k=10$ are somewhat similar (and small) which explains the reason why  the curves for the probabilities of DCE are close.
%---------
Finally, the bottom panel of Figure \ref{fig:sim:prob:DCE} reports the overall probability of DCE in Equation
\eqref{eq:prob:DCE}, which averages $p^{DC}_0(k_0,n)$ and $p^{DC}_1(k_1,n)$ with $P(H_0)=P(H_1)=0.5$ following Equation \eqref{eq:prob:H0}.

\begin{figure}%[b]
\centering
\includegraphics[scale=0.55]{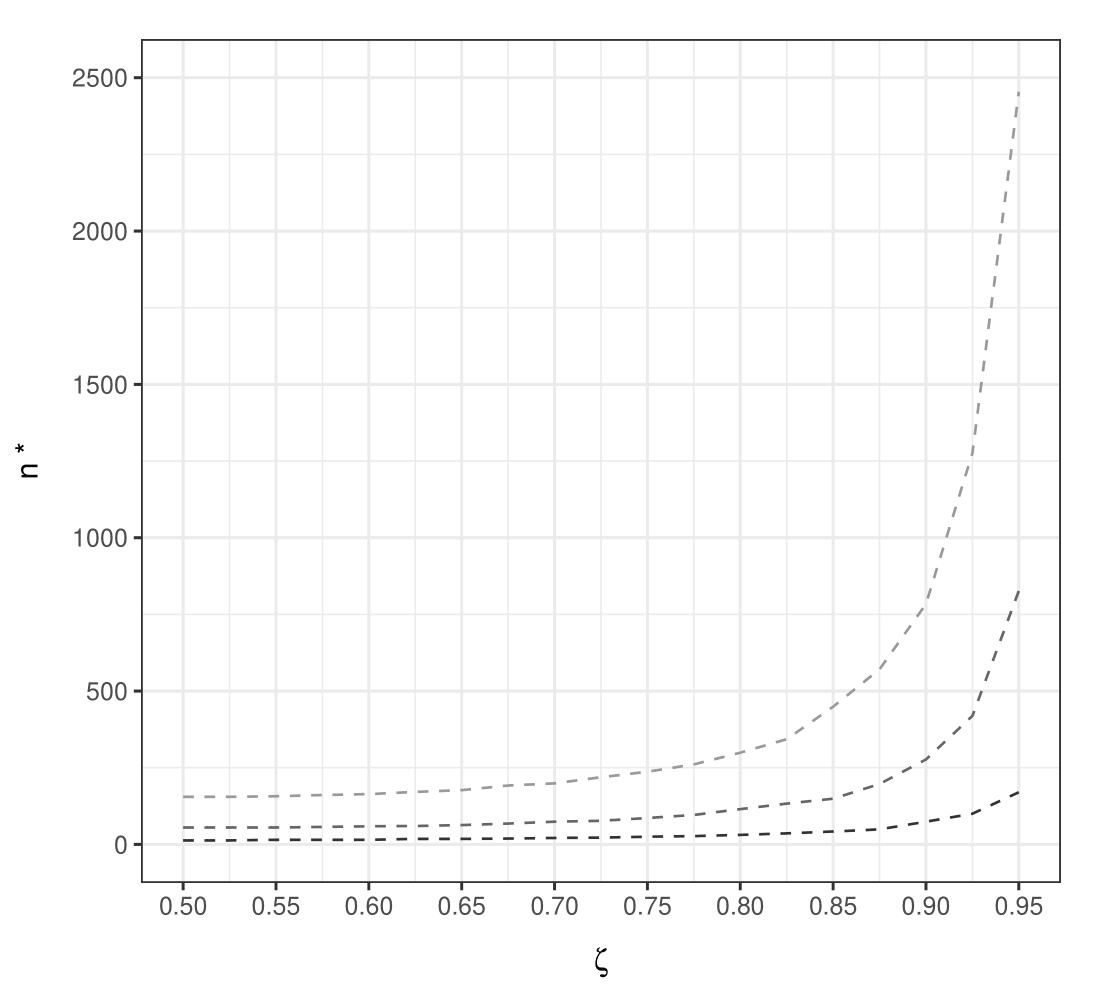}
\caption{Two-node chain component: Optimal sample size $n_{uv}^*=n^*$ as a function of threshold $\zeta\in[0.50,0.95]$ for different values of $k\in\{3,6,10\}$ (from dark to light grey) and $k_0=k_1=k$.}
\label{fig:sim:n:star}
\end{figure}

We now move to SSD and obtain the optimal sample size $n_{uv}^*$ for an intervention on $u$ based on \eqref{eq:optim:sample:size}.
The latter quantity is computed for each value of the BF threshold $k\in\{3,6,10\}$
and for distinct thresholds for the probability of DCE $\zeta\in[0.5,\dots,0.95]$.
Results are summarized in Figure \ref{fig:sim:n:star} which reports the behavior of $n_{uv}^*$ as a function of $\zeta$ for the three increasing levels of $k$ (from dark to light gray).
Clearly, the optimal sample size required for DCE increases  with the threshold $\zeta$. The behavior of the three curves as $k$ varies is similar; however it becomes much steeper beyond $\zeta=0.85$  for $k=10$ (the cutoff which separates moderate from strong evidence).
As an example, if we fix $\zeta=0.8$, we obtain an optimal sample size $n_{uv}^*\simeq 50$ for $k=3$, and this value triples  when $k=6$ and reaches $n_{uv}^*\simeq300$ for $k=10$. The latter sample size would instead guarantee a probability of DCE higher than $95\%$ when $k=3$.

\subsection{Riboflavin data}

%From Maathuis et al.
In this section we apply our strategy for sample size determination to a data set about riboflavin
(vitamin B2) production by \textit{Bacillus subtilis}.
The dataset is publicly available within the \texttt{R} package \citep{R:core:team} \texttt{hdi}
and includes $q=4089$ variables, namely the logarithm of the riboflavin production
rate  and the  log-expression level of $4088$ genes that cover essentially the whole genome of
\textit{Bacillus subtilis}.
The sample size is $N=71$.
This observational dataset was analyzed by \citet{Maat:etal:2009} to infer causal effects on the riboflavin production rate due to single gene manipulations.
To this end the authors first estimate a CPDAG using the PC algorithm \citep{Spir:Glym:Sche:2000}.
Then, using do-calculus theory, they provide an estimate of the causal effect on the riboflavin rate following an hypothetical intervention on each of the $4088$ nodes.
Each causal effect is not unique in general because it depends on the specific set of parents of the node (also called \textit{adjustment set}),  and ultimately  on the underlying DAG structure; see also \citet[Algorithm 1]{Maat:etal:2009}.
Since typically many DAGs are compatible with the input CPDAG, a collection of possible causal effects (and eventually the corresponding average) is finally provided by their procedure.

\begin{figure}%[b]
\centering
\includegraphics[scale=0.6]{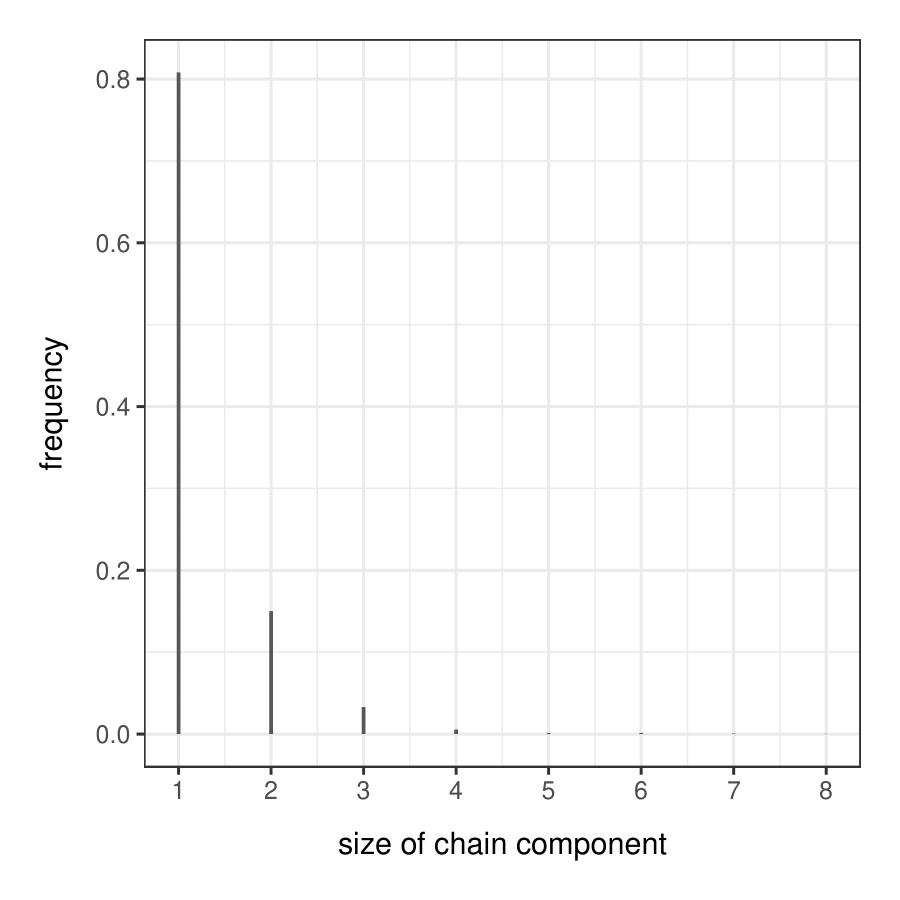}
\caption{Riboflavin data: Distribution of the size of chain components in the estimated CPDAG.}
\label{fig:riboflavin:size:cc}
\end{figure}

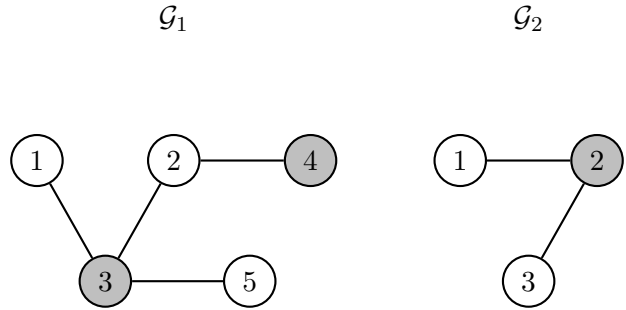
\begin{figure}
\begin{center}
	\begin{tabular}{ccccc}
		$\G_1$ & & & $\G_2$ \\
		\\ \\
		\begin{tikzpicture}
			\vspace{0.5cm}
			\begin{scope}[every node/.style={circle,thick,draw,fill=lightgray}]
				\node (4) at (0.9,-1.6) {3};
				\node (3) at (3.6,0) {4};
			\end{scope}
			\begin{scope}[every node/.style={circle,thick,draw}]
				\node (1) at (0,0) {1};
				\node (2) at (1.8,0) {2};
				\node (5) at (2.8,-1.6) {5};
			\end{scope}
			\begin{scope}[>={Stealth[black]},
				every node/.style={fill=white,circle},
				every edge/.style={draw=black, thick}]
				\path [-] (1) edge (4);
				\path [-] (4) edge (2);
				\path [-] (2) edge (3);
				\path [-] (4) edge (5);
			\end{scope}
		\end{tikzpicture}
		& & &
		\begin{tikzpicture}
			\vspace{0.5cm}
			\begin{scope}[every node/.style={circle,thick,draw,fill=lightgray}]
				\node (2) at (1.8,0) {2};
			\end{scope}
			\begin{scope}[every node/.style={circle,thick,draw}]
				\node (1) at (0,0) {1};
				\node (3) at (0.9,-1.6) {3};
			\end{scope}
			\begin{scope}[>={Stealth[black]},
				every node/.style={fill=white,circle},
				every edge/.style={draw=black, thick}]
				\path [-] (1) edge (2);
				\path [-] (2) edge (3);
			\end{scope}
		\end{tikzpicture}
	\end{tabular}
\end{center}
\caption{Riboflavin data: Two chain-component sub-graphs, $\G_1, \G_2$, with grey dots representing BOS manipulated variables.}
\label{fig:riboflavin:chain:comp}
\end{figure}

We take a different course of action and apply our method to estimate the optimal sequence of manipulated variables and corresponding sample sizes as in Section \ref{sec:Bayes:SSD:active:learning}.
We start from an input CPDAG $\mathcal{E}$ estimated using the PC algorithm with the tuning  parameter $\alpha$ set at level $0.01$.
We then fix the threshold for the probability of DCE $\zeta=0.8$, while $k_0=k_1=6$.
Recall now that our objective is SSD for an intervention needed to orient those edges which are undirected in the input CPDAG.
Since undirected edges can only occur between (two ore more) nodes belonging to the same chain component,
we focus on those chain components whose size is larger than one.
Figure \ref{fig:riboflavin:size:cc} summarizes the distribution of the size of chain components in the input CPDAG $\E$.
Most of these (about $80\%$) have size equal to one, in which case there are no edges whose orientation needs to be determined.
We then implement our method on each of the remaining chain components separately.
As an example, Figure \ref{fig:riboflavin:chain:comp} reports two chain component sub-graphs $\G_1, \G_2$.
%(with nodes labelled as $\{1,\dots,T\}$ in both cases)
with the corresponding Best size Optimal Sequences of manipulated variables (BOS) represented as grey nodes.
\black
For $\G_1$ there are actually  two optimal sequences  according to Definition \ref{def:optimal:he:geng}, namely  $S_1=(2,3)$ and $S_2=(3,4)$, as we report  in Table \ref{tab:riboflavin:optimal:sss} with the corresponding optimal sample sizes computed as in
\eqref{alg:optimal:sequence:sample:size}. Clearly $S_2$ is the BOS because the \emph{total} sample size  under $S_2$ is smaller than under $S_1$.
On the other hand, there is only one optimal sequence for $\G_2$, whose corresponding optimal sample size is $130$ (see again Table \ref{tab:riboflavin:optimal:sss}).

As an overall summary, we also report in Table \ref{tab:riboflavin:summary}, for each size of the chain components of $\E$ and size of the  BOS, the corresponding average total sample size (Average $N^*$)   and average sample size \textit{per} intervention (Average $n^*$), with the average computed across sequences of manipulated variables.
It appears that while in general the number of interventions needed for edge orientation (Size of sequence $S^*$) increases with the size of the chain component,
the total sample size is not typically higher for larger (chain components and) sizes of $S^*$; compare for instance $s=4$ and $s=5$.
In addition, the average sample size \textit{per} intervention, for each given value of $s$ for which different sizes of $S^*$ are observed, is smaller for those sequences of manipulated variables having larger sizes.
Accordingly, while more interventions are needed to orient edges in chain components of larger dimension, the (optimal) number of interventional data (sample size) required by each intervention
%to reach the pre-determined level of DCE
is in general smaller. This  suggests the existence of a trade-off between the number of manipulated variables and the optimal sample size \textit{per} intervention.

\begin{table}
\centering
\caption{Riboflavin data: Optimal sequences of manipulated nodes and corresponding sample sizes for each of the two chain-component sub-graphs, $\G_1,\G_2$ in Figure \ref{fig:riboflavin:chain:comp}.}
\label{tab:riboflavin:optimal:sss}
\begin{tabular}{ccc}
	\hline
	& Optimal sequence of nodes & Optimal sample size \\
	\hline
	\multirow{2}{*}{$\G_1$}
	& $S_1=(2,3)$ & $\boldsymbol{n}^{*(1)}=(28,88)$ \\
	& $S_2=(3,4)$ & $\boldsymbol{n}^{*(2)}=(4,86)$ \\
	\hline
	\multirow{1}{*}{$\G_2$}
	& $S_1=2$ & $\boldsymbol{n}^{*(1)}=130$ \\
	\hline
\end{tabular}
\end{table}

\begin{table*}
\centering
\caption{Riboflavin data: average total sample size (Average $N^*$) and average sample size \textit{per} manipulated variable (Average $n^*$) cross-classified by size of chain component of the input CPDAG and size of the sequence of manipulated variables  $S^*$.}
\label{tab:riboflavin:summary}

\begin{tabular}{ccccccccccc}
	\hline
	Size of chain component && \multicolumn{1}{c}{$s=2$} && \multicolumn{2}{c}{$s=3$} && \multicolumn{2}{c}{$s=4$} && \multicolumn{1}{c}{$s=5$} \\
	\hline
	Size of sequence $S^*$ && 1 && 1 & 2 && 1 & 2 && 2 \\
	\hline
	Average $N^*$ && 48.3 && 131.4 & 100.0 && 241.3 & 248.4 && 81.0  \\
	Average $n^*$
	&& 48.3 && 131.4 & 50.0 && 241.3 & 124.2 && 40.5 \\
	\hline
\end{tabular}

\vspace{0.5cm}

\begin{tabular}{cccccc}
	\hline
	\multicolumn{2}{c}{$s=6$} && \multicolumn{1}{c}{$s=7$} && \multicolumn{1}{c}{$s=8$} \\
	\hline
	 2 & 3 && 3 && 4 \\
	\hline
	 67.0 & 37.0 && 65.5 && 117.0 \\
	 33.5 & 12.3 && 21.8 && 29.3 \\
	\hline
\end{tabular}
\end{table*}

%491  108   18    4    4    2    1

\section{Discussion}
\label{sec:discussion}

Observational data cannot distinguish in general among different DAG-models representing the same conditional independence assertions.
This is a serious drawback for causal inference which is predicated on a given DAG representing the data generating process, as required by  do-calculus theory.
%-------------
Intervention experiments, leading to the collection of interventional data, can greatly improve the structure learning process.
So far most works in active learning have concentrated
on efficient algorithms to select target variables to intervene upon in order to guarantee  identification of the underlying causal DAG, starting from a Markov equivalence class of DAGs.
Interventions on variables help decide how to orient undirected edges which are present in the CPDAG representative of the equivalence class. However the actual decision is based on sampling data,  as in the independence test between a target variable and one of its neighbors.
%-----------
Active learning involves therefore two aspects,  one is algorithmic and uses graph-based notions  for the selection of the target variables, and the other one is statistical and uses samples of interventional data, possibly coupled with  external information,  which may include previously collected observational data or substantive domain knowledge.
%Both are crucial for an effective implementation; yet the statistical side has been rather neglected so far.
In this context,  a question which has so far been neglected
is the determination of the sample size of the  interventional data
required to achieve desirable inferential properties.
This paper addresses this issue
with regard to the problem of edge orientation,
which is framed as
a test of hypothesis between two competing DAG structures. Specifically, we use the Bayes factor as a measure of evidence,  and for a given  sequence of optimally specified intervention variables, we determine the corresponding  collection of sample sizes which will produce decisive and correct evidence in favor of the true causal-model hypothesis at each intervention.
%%%%%%%%%%%%%%

Our method takes as  input an equivalence class of DAGs,  equivalently
its representative CPDAG, which  typically has been estimated from an observational dataset.
It does not accommodate for estimation uncertainty, as most active learning methods do.
%------------
In principle the posterior distribution over the space of Markov equivalence classes, see  e.g.  \citet{Castelletti:et:al:2018}, could  be of some help to evaluate the strength of the evidence in favor of the chosen CPDAG, possibly the highest posterior probability model.
Operationally,  however,
one would still need
a \emph{single} sequence of variables to manipulate in order to perform sample size determination (SSD), and
it is far from clear how
standard Bayesian Model Averaging techniques
could be used
in a fruitful way.

%questo quanto c'era in Castelletti e Cons 2020 JRSSA

%We see two main limitations  in this approach. First, it requires as input  an  equivalence class estimated through observational data. Unfortunately there is no indication of whether alternative equivalence classes might be also likely  as starting points, because no  measure of inferential uncertainty is provided. Accordingly we see a robustness issue here.

%Also from their abstract:
%We present a Bayesian active learning procedure for Gaussian DAGs which requires
%no subjective specification on the side of the user, explicitly takes into account the uncertainty
%on the space of equivalence classes (through the posterior distribution) and sequentially proposes
%the choice of the optimal intervention variable

Our method for SSD is based on a sequence of manipulated variables  arising from a \emph{batch} intervention experiment.
%The latter guarantees that all DAGs within the input Markov equivalence class
%are identifiable once the entire interventional dataset is collected.
A \emph{sequential} approach would instead proceed by choosing the intervention nodes one at a time and collecting new data after each intervention \citep{Stefan:et:al:2022,He:etal:2008}.
In this way,
the optimal sample size associated with a target node could be computed, at each stage,  using all the samples collected up to that step,
thus
increasing the amount of information used for prediction at the design level.
%----------------
Another advantage of the sequential method
is to alleviate
the danger inherent in the choice of the starting Markov equivalence class, namely that  the true DAG could be outside the class. \citet{Hauser:Buehlmann:2014} investigate this aspect and illustrate by simulation that methods that do take into account observational as well as interventional data show  a better performance in recovering the true data-generating DAG.
We observe that our method
could be tailored to a sequential setup by
incorporating in
the predictive distribution of the BF
not only the initial  observational data but also the newly  interventional observations collected at each step.
%-------

%\purple to say here or in a Discussion

%we do not treat SSD according to  sequential approach, e.g. cite for recent references  Stefan et al and also  Schoenbrodt e Wagenmakers. In senso lato tuttavia un po' di sequenziale c'è in He and Geng nella scelta degli interventi

% \purple
%
%[3. Robustness]
%
%\green lo lasciamo? in tal caso andrebbe letto De Santis 2006 per dare un minimo di contributo
%\purple
%another aspect that we do not consider is robustness of SSD to model mis-specificationvedere
%(vedere prima De Santis F. (2006). Sample size determination for robust Bayesian analysis. Journal
%of the American Statistical Association, vol. 101, n. 473, 278-291.
%Invece non mi pare che quel che dicono Wang and Gelfand sia utile per noi in quanto loro di occupano di discriminare ad es una normale da una t con 2 gr lib)
%
%

%[4. hard and soft;  latent]

In our procedure experimental data are generated under hard  interventions and in the absence of latent variables.
%[hard soft]
%pezzo Yang Katkoff e Uhler
%This setting is common in biology, where gene regulatory networks can be intervened on using chemical reagents or gene deletions. Hauser & Buhlmann (2012) previously characterized the identifiability of causal DAGs under
Under hard (or perfect) interventions, dependencies between targeted variables and their direct causes are removed. This assumption may not hold in some settings where dependencies can only be altered without being fully deleted.
An instance is genomic medicine, where gene manipulation through repression or activation of selected genes is performed to better understand the complex functioning of the pathway.
% Questo da Dominguez et al., 2016
Intervention experiments for gene regulation are meant to be perfect
but in practice may not be uniformly successful across a cell population,
in which case the dependence between manipulated genes and their direct causes in the network is only weakened but maintained.
Identifiabilty of causal DAGs from soft interventions is investigated from a theoretical perspective by \citet{Yang:Uhler:2018} who propose a consistent algorithm for DAG structure learning.

%Complex and dynamic transcription regulation of multiple genes and their pathways drives many essential cellular activities, including genome replication and repair, cell division and differentiation, and disease progression and inheritance. Understanding the complex functions of a gene network requires the ability to precisely manipulate and perturb expression of the desired genes by repression or activation.

%However, only considering perfect interventions is restrictive: in practice, many interventions are non-perfect (or
%soft) and modify the causal relations between targeted variables and their direct causes without eliminating them (Eberhardt et al., 2005). In genomics, for example, interventions
%such as RNA interference or CRISPR-mediated gene activation often have only modest effects on gene suppression
%and activation respectively (Dominguez et al., 2016). Even
%interventions meant to be perfect, such as CRISPR/Cas9-
%mediated gene deletions, may not be uniformly successful
%across a cell population (Dixit et al., 2016). Although nonperfect interventions may be considered inefficient from
%an engineering perspective, they may still provide valuable
%information about regulatory networks. The identifiability
%of causal DAGs in this setting needs to be formally analyzed
%to develop maximally effective algorithms for learning from
%these types of interventions.
%------------
%[latent]
%------------
Finally, our theoretical framework is that of
a causally sufficient system with no latent confounders, selection
bias, or feedback. Some works on structural learning try to alleviate  these limitations;
see for instance \citet{Frot:et:al:2019} and \citet{squi:etal:arxiv:2020}.

\black

%%%%%%%%%%%%%%%%%%%%%%%%%%%%%%%%%%%%%%%%%%%%%%
%% Example with multiple Appendixes:        %%
%%%%%%%%%%%%%%%%%%%%%%%%%%%%%%%%%%%%%%%%%%%%%%

\begin{appendix}
\label{sec:appendix}

\section{Priors for DAG model comparison and Bayes Factor computation}\label{appA}

Our elicitation scheme for parameter priors under a general Gaussian DAG model is based on the procedure introduced by \citet{Geig:Heck:2002} (G\&H).
This is used both to construct the analysis prior  and the design prior. The former is needed to obtain the Bayes factor, whose predictive distribution is generated under the latter.

\subsection{General assumptions}

The method of G\&H
is based on a set of assumptions which
drastically simplifies the elicitation of priors; additionally
it ensures
\emph{compatibility} of priors across DAG models,
so that
%a notion that is especially critical for  DAG-graphical models, as we clarify below.
%\citet{Geig:Heck:2002} (G\&H) propose a  method to construct parameter priors for the comparison of DAG-models which ensures identical marginal likelihoods for
DAGs belonging to the same equivalence class score the same marginal likelihood.
This feature is important when DAG model comparison is based on observational data, because the latter cannot distinguish in general among Markov equivalent DAGs.
This however is no longer the case when  interventional data are also employed, as we do in Section \ref{sec:analysis:prior:BF}.

%\st{We emphasize that while the same formula is adopted to compute the BF of $H_0$ against $H_1$ in} Section \ref{sec:analysis:prior:BF}, \st{the two model hypotheses under comparison refer to post-intervention distributions of $\D_0$ and $\D_1$ which are \emph{not} Markov equivalent anymore because of the intervention on node $u$.}

The method assumes some regularity conditions on the likelihood, namely
\textit{complete model equivalence, regularity and likelihood modularity}
which are satisfied by any Gaussian model.
In addition, two assumptions on the prior distributions are introduced.
%The distinctive feature  of their approach
%regards however the construction of the prior. This can be split into two steps.
The first assumption
(\textit{prior modularity}) states that, for any two distinct DAG models with the \emph{same} set of parents for vertex $j$, the  prior for the node-parameter $\btheta_j$ must be the same under both models.
%\be
%p(\btheta_j \g \D_h) = p(\btheta_j \g \D_k)
%\ee
%for any pair of distinct DAGs $\D_h$ and $\D_k$ such that $\pa_{\D_h}(j)=\pa_{\D_k}(j)$.
Moreover, the second assumption
(\textit{global parameter independence}) states that for every DAG model $\D$, the parameters $\{\btheta_j; \, j=1, \ldots, q\}$ should be \textit{a priori} independent, that is
\be
p(\btheta\g\D)=\prod_{j=1}^{q}p(\btheta_j\g\D).
\ee
Based on these assumptions, Theorem 1 of \citet{Geig:Heck:2002} shows that the parameter priors of \emph{all} DAG models are completely determined by a \emph{unique}  prior on the parameter of \emph{any} of the (equivalent) complete DAGs.

Specifically, in the zero-mean Gaussian framework, all priors across DAG models
can be shown to be driven by a \emph{single} Wishart distribution on an \emph{unconstrained} precision matrix.
Most importantly, a direct consequence of the method is that each marginal data distribution in Equation \eqref{eq:BF:01:indep:vs:dep}
corresponds to the marginal data distribution computed under any complete DAG model; see  next section for more details.

\subsection{Marginal data distributions and Bayes Factor}

Consider a multivariate Gaussian model of the form
\ben
\begin{aligned}
	\label{eq:gaussian:model}
	\bx_1,\dots,\bx_n \g \bOmega &\overset{\textnormal{iid}}{\sim}
	\N_T\left(\bzero, \bOmega^{-1}\right)\\
	%\nonumber
	\bOmega&\sim\W_T(a,\bU),
\end{aligned}
\een
where $\W_T(a,\bU)$ denotes a Wishart distribution having expectation $a\bU^{-1}$ and $a>T-1$.
Let also $\bS=\sum_{h=1}^n\bx_h\bx_h^\top$.
The marginal data distribution restricted to variables in $A\subseteq\{1,\dots,T\}$ is given by
\ben
\label{eq:marg:like:general}
\nonumber m(\bX_A)=\frac{\prod_{j=1}^{|A|}\Gamma\left(\frac{a-|\bar{A}|+n+1-j}{2}\right)}
{\prod_{j=1}^{|A|}\Gamma\left(\frac{a-|\bar{A}|+1-j}{2}\right)}
\frac{\left|\bU_{A,A}\right|^{\frac{a-|\bar{A}|}{2}}}
{\left|\bU_{A,A}+\bS_{A,A}\right|^{\frac{a-|\bar{A}|+n}{2}}},\\
\een
where $\bU_{A,A}$ denotes the sub-matrix of $\bU$ with rows and columns indexed by $A$ and $\bar{A}=\{1,\dots,T\}\setminus A$;
see for instance \citet[Equation 12]{Consonni:LaRocca:2012}.
Moreover for simplicity in this section we omit superscript $n$ from data matrices.
%---------
Under the Gaussian setting of Section \ref{sec:BF:predictive:Gaussian},
the BF in Equation \eqref{eq:BF:01:indep:vs:dep}
can be evaluated using
the marginal likelihood \eqref{eq:marg:like:general} for $A=u$, $A=v$ and $A=\{u,v\}$.
We thus obtain
\ben
\label{eq:marg:like:u}
\nonumber m(\bX_u)=\frac{\Gamma\left(\frac{a-(T-1)+n}{2}\right)}
{\Gamma\left(\frac{a-(T-1)}{2}\right)}
\frac{\left|\bU_{u,u}\right|^{\frac{a-(T-1)}{2}}}
{\left|\bU_{u,u}+\bS_{u,u}\right|^{\frac{a-(T-1)+n}{2}}}\\
\een
and similarly for $A=v$, while for $A=\{u,v\}$,
\ben
\label{eq:marg:like:uv}
\nonumber m(\bX_{u,v})&=&
\frac{\Gamma\left(\frac{a-(T-2)+n}{2}\right)}
{\Gamma\left(\frac{a-(T-2)}{2}\right)}
\frac{\Gamma\left(\frac{a-(T-2)+n-1}{2}\right)}
{\Gamma\left(\frac{a-(T-2)-1}{2}\right)}\\
&\cdot&
\frac{\left|\bU_{\{u,v\},\{u,v\}}\right|^{\frac{a-(T-1)}{2}}}
{\left|\bU_{\{u,v\},\{u,v\}}+\bS_{\{u,v\},\{u,v\}}\right|^{\frac{a-(T-1)+n}{2}}}.
\een
\noindent
Therefore, the BF in \eqref{eq:BF:01:indep:vs:dep} reduces to
\ben
\label{eq:BF:0:1:subj:formula}
\nonumber
\textnormal{BF}_{01}^n &=&
\frac{\Gamma\left(\frac{a-(T-1)+n}{2}\right)}
{\Gamma\left(\frac{a-(T-1)}{2}\right)}
\cdot
\frac{\Gamma\left(\frac{a-(T-2)}{2}\right)}
{\Gamma\left(\frac{a-(T-2)+n}{2}\right)} \\
&\cdot&
\frac{\left[\bU_{u,u}\bU_{v,v}\right]^{\frac{a-(T-1)}{2}}}
{\left|\bU_{\{u,v\},\{u,v\}}\right|^{\frac{a-(T-2)}{2}}}
\cdot
\frac{\left|\bU_{\{u,v\},\{u,v\}}+\bS_{\{u,v\},\{u,v\}}\right|^{\frac{a-(T-2)+n}{2}}}
{\left[(\bU_{u,u}+\bS_{u,u})(\bU_{v,v}+\bS_{v,v})\right]^{\frac{a-(T-1)+n}{2}}}.
\een

\vspace{0.5cm}

So far  results were obtained under a subjective prior on $\bOmega$.
We now consider an \emph{objective} framework based on the notion of \textit{Fractional Bayes Factor} (FBF) \citep{Ohagan:1995}.
Specifically, we start from the default objective prior
\ben
\label{eq:default:prior}
p^D(\bOmega) \propto |\bOmega|^{\frac{a_{\Omega}-T-1}{2}}.
\een
%-----------
Let now $\bar{\bS}=\frac{1}{n}\bS$.
The (data dependent) fractional prior on $\bOmega$ is defined as
\be
p^F(\bOmega)\propto \left\{p(\bX\g\bOmega)\right\}^{b}p^D(\bOmega),
\ee
where $b\in(0,1)$ is typically chosen as the smallest value s.t. the fractional prior is proper.
After some calculations we obtain
\be
\bOmega \sim \W_{T}(a_{\Omega}+n_0,n_0\bar{\bS}),
\ee
where $n_0=bn$, which is proper provided  $a_{\Omega}+n_0>T-1$;
see \citet{Consonni:LaRocca:2012} for full details.
\black
Also, the posterior distribution of $\bOmega$ is
\be
p(\bOmega\g \bX)&\propto& \left\{p(\bX\g\bOmega)\right\}^{1-b}p^F(\bOmega) \\
\brown &=& \left\{p(\bX\g\bOmega)\right\}p^D(\bOmega).
\ee
%--------
\black
\noindent
The FBF is obtained by specializing \eqref{eq:BF:0:1:subj:formula}
with
\ben
\nonumber
a \mapsto a_{\Omega}+n_0, \quad
n \mapsto n - n_0, \\
\nonumber
\bU \mapsto \frac{n_0}{n} \bS, \quad
%= \frac{n_0}{n}\bS
\bS \mapsto \frac{n-n_0}{n} \bS,
\een
which after some calculations leads to
\ben
\label{eq:BF:0:1:obj:formula}
\nonumber
\textnormal{BF}_{01}^n &=&
\frac{\Gamma\left(\frac{a_{\Omega}-(T-1)+n}{2}\right)}
{\Gamma\left(\frac{a_{\Omega}+n_0-(T-1)}{2}\right)}
\cdot
\frac{\Gamma\left(\frac{a_{\Omega}+n_0-(T-2)}{2}\right)}
{\Gamma\left(\frac{a_{\Omega}-(T-2)+n}{2}\right)}
\cdot
\left(\frac{n_0}{n}\right)^{-1}
\left[\frac{\left|\bS_{\{u,v\},\{u,v\}}\right|}
{\bS_{u,u}\bS_{v,v}}\right]^{\frac{n-n_0}{2}}.
\nonumber
\een
Now notice that
\be
\left|\bS_{\{u,v\},\{u,v\}}\right|
=\sum_{h=1}^nx_{h,u}^2\sum_{h=1}^nx_{h,v}^2
-\left(\sum_{h=1}^nx_{h,u}x_{h,v}\right)^2
\ee
and
\be
\bS_{u,u}=\sum_{h=1}^nx_{h,u}^2, \quad
\bS_{v,v}=\sum_{h=1}^nx_{h,v}^2.
\ee
Therefore, we can write
\ben
\begin{aligned}
	\frac{\left|\bS_{\{u,v\},\{u,v\}}\right|}
	{\bS_{u,u}\bS_{v,v}}
	&=
	1- \frac{\left(\sum_{h=1}^nx_{h,u}x_{h,v}\right)^2}
	{\sum_{h=1}^nx_{h,u}^2\sum_{h=1}^nx_{h,v}^2}\\
	&=
	1-\left(r_{u,v}^n\right)^2
\end{aligned}
\een
where $r_{u,v}^n$ denotes the sample correlation coefficient between $X_u$ and $X_v$.
In the sequel  we choose  $a_{\Omega}=T-1$ so that the prior
is proper  even with a training sample size $n_0$  equal to one, and
%$p(\bOmega) \propto |\bOmega|^{-1}$.
\black
we obtain
\ben
\begin{aligned}
	\textnormal{BF}_{01}^n &=
	\frac{1}{\sqrt{\pi}}\frac{\Gamma\left(\frac{n}{2}\right)}{\Gamma\left(\frac{n+1}{2}\right)}n
	\left[1-(r_{uv}^n)^2\right]^{\frac{n-1}{2}}.
	%\\
	%&=
	%A(n)\left[1-(r_{uv}^n)^2\right]^{\frac{n-1}{2}}.
\end{aligned}
\een

\vspace{0.2cm}

\subsection{Posterior distribution of DAG model parameters}

The design prior for $(\bL_{u,v},\bD_{v,v})$
that we adopt in Section \ref{subsec:posterior predcitive}
corresponds to the
posterior $p(\bL_{u,v},\bD_{v,v}\g\bZ,\D_1)$.
%\be
%\bL_{u,v}=-\left(\bSigma_{\{u,v\},\{u,v\}}\right)^{-1}\bSigma_{u,v},
%\quad
%\bD_{v,v} = \bSigma_{v\g u}.
%\ee
The latter can be recovered from the posterior on $\bOmega=\bSigma^{-1}$, the (unconstrained) precision matrix of a complete DAG,
following the procedure of G\&H, which we detail below.

Let $\D$ be an arbitrary DAG %and assume a parent ordering of its nodes.
and let  $\prec j \succ \, =\pa(j)$,  and  $\prec j\,]=\pa(j)\times j$.
Consider the (Cholesky) re-parameterization $\bOmega\mapsto(\bL,\bD)$ where, for $j \in \{1,\dots,q\}$,
\be
\bD_{jj}=\bSigma_{jj\g\pa_{\D}(j)}, \quad \bL_{ \prec j\,]}=-\bSigma^{-1}_{ \prec j\, \succ}\bSigma_{ \prec j\,]}
\ee
For each node $j \in \{1,\dots,q\}$, let  $\big\{\bD_{jj},\bL_{\prec j \, ]}\big\}$ be the  parameters associated to node  $j$, and identify a complete DAG $\D^{C(j)}$ such that $\pa_{\D^{C(j)}}(j)=\pa_{\D}(j)$.
% under $\bOmega_{\D}$, $\bOmega_{\D} \in \mathcal{P}_{\D}$.
%where for simplicity of notation we omit from $\bOmega$ the dependence on the ordering of the variables under $C(j)$.
Let
$\big\{\bD_{j j}^{C(j)},\bL_{\prec j \, ]}^{C(j)}\big\}$ be the  parameters of node $j$ under the complete DAG $\D^{C(j)}$.
We then assign to $\left\{\bD_{jj}, \bL_{\prec j \, ]}\right\}$ the same prior of $\big\{\bD_{jj}^{C(j)},\bL_{\prec j \, ]}^{C(j)}\big\}$.
However, because our interest is in obtaining the posterior of DAG  parameters  $(\bD,\bL)$,
we can compute first  the posterior on the unconstrained $\bOmega$, which by conjugacy is still Wishart,
and then recover the posterior on $(\bD,\bL)$.
%-----------
%\purple obs: noi facciamo nell'ordine: posterior di $\bOmega$, posterior di $\bSigma_{\{u,v\},\{u,v\}}$, quindi posterior di $\bL_{u,v}, \bD_{v,v}$
%---------

%----------------
Consider a random sample of size $N$, $\bz_1, \dots, \bz_N$, with $\bz_i=(x_{i,1},\dots,x_{i,q})^\top$ and
$\bz_{i}\g \bOmega \overset{\textnormal{iid}}{\sim} \N_q(\boldsymbol{0},\bOmega^{-1})$, $i = 1,\dots,N$, where $\bOmega$ is unconstrained.
Let also $\bZ$ be the $(N,q)$ data matrix, obtained by row-binding  the individual $\bz_i^\top$'s.
The posterior distribution of $\bOmega$ computed under the default prior
\eqref{eq:default:prior}
is given by
\be
\bOmega \g \bZ \sim \W_{T}(a_{\Omega}+n,\bZ^\top\bZ).
\ee
To obtain draws from the posterior of $(\bL_{u,v},\bD_{v,v})$, set
$
\left(\bSigma_{\{u,v\},\{u,v\}}\right)^{-1}
%= \bOmega_{\{u,v\},\{u,v\}\g\tau\setminus\{u,v\}}
:= \boldsymbol{Q}_{u,v}
$,
and  using properties of the Wishart distribution deduce
\ben
\label{eq:dist:Quv}
\quad \quad
\boldsymbol{Q}_{\{u,v\}}
\sim\W_2\left(a_{\Omega}+N-(T-2),\bS_{\{u,v\},\{u,v\}}\right);
\een
see \citet[Thm 5.1.4]{Press:1982} and \citet[Section 2.1]{Consonni:LaRocca:2012}.
Consider now a draw from \eqref{eq:dist:Quv} and compute
$\bSigma_{\{u,v\},\{u,v\}}$.
Finally recover
\be
\bL_{u,v}=-\left(\bSigma_{\{u,v\},\{u,v\}}\right)^{-1}\bSigma_{u,v},
\quad
\bD_{v,v} = \bSigma_{v\g u},
\ee
where $\bSigma_{v\g u}=\bSigma_{v,v}-(\bSigma_{v,u})^2\bSigma_{u,u}^{-1}$.

\end{appendix}

% BibTeX users please use one of
\bibliographystyle{biometrika}       % APS-like style for physics
\bibliography{biblio_SS}   	         % name your BibTeX data base

\begin{thebibliography}{71}
\expandafter\ifx\csname natexlab\endcsname\relax\def\natexlab#1{#1}\fi
\expandafter\ifx\csname url\endcsname\relax
  \def\url#1{\texttt{#1}}\fi
\expandafter\ifx\csname urlprefix\endcsname\relax\def\urlprefix{URL }\fi
\providecommand{\eprint}[2][]{\url{#2}}

\bibitem[{Adcock(1997)}]{Adcock:1997}
\textsc{Adcock, C.~J.} (1997).
\newblock Sample size determination: A review.
\newblock \textit{J. R. Stat. Soc. D. Stat.} 46 261--283.

\bibitem[{Andersson et~al.(1997)Andersson, Madigan \& Perlman}]{Ande:etal:1997}
\textsc{Andersson, S.~A.}, \textsc{Madigan, D.} \& \textsc{Perlman, M.~D.}
  (1997).
\newblock {A characterization of Markov equivalence classes for acyclic
  digraphs}.
\newblock \textit{Ann. Statist.} 25 505--541.

\bibitem[{Andersson et~al.(2001)Andersson, Madigan \&
  Perlman}]{Andersson:et:al:2001}
\textsc{Andersson, S.~A.}, \textsc{Madigan, D.} \& \textsc{Perlman, M.~D.}
  (2001).
\newblock Alternative {M}arkov properties for chain graphs.
\newblock \textit{Scand. J. Stat.} 28 33--85.

\bibitem[{Castelletti \& Consonni(2020)}]{Castelletti:Consonni:2020}
\textsc{Castelletti, F.} \& \textsc{Consonni, G.} (2020).
\newblock Discovering causal structures in bayesian gaussian directed acyclic
  graph models.
\newblock \textit{J. R. Stat. Soc. Ser. A. Stat. Soc.} 183 1727--1745.

\bibitem[{Castelletti \& Consonni(2021)}]{Castelletti:Consonni:Biom:2021}
\textsc{Castelletti, F.} \& \textsc{Consonni, G.} (2021).
\newblock {Bayesian inference of causal effects from observational data in
  Gaussian graphical models}.
\newblock \textit{Biometrics} 77 136--149.

\bibitem[{Castelletti et~al.(2018)Castelletti, Consonni, Della~Vedova \&
  Peluso}]{Castelletti:et:al:2018}
\textsc{Castelletti, F.}, \textsc{Consonni, G.}, \textsc{Della~Vedova, M.} \&
  \textsc{Peluso, S.} (2018).
\newblock {Learning Markov equivalence classes of directed acyclic graphs: An
  objective Bayes approach}.
\newblock \textit{Bayesian Anal.} 13 1235--1260.

\bibitem[{Castelo \& Perlman(2004)}]{Cast:Perl:2004}
\textsc{Castelo, R.} \& \textsc{Perlman, M.~D.} (2004).
\newblock Learning essential graph {M}arkov models from data.
\newblock In \textit{Advances in {B}ayesian networks}, vol. 146 of
  \textit{Stud. Fuzziness Soft Comput.} Springer, Berlin, 255--269.

\bibitem[{Chaloner \& Verdinelli(1995)}]{Chaloner:Verdinelli:1995}
\textsc{Chaloner, K.} \& \textsc{Verdinelli, I.} (1995).
\newblock {Bayesian experimental design: A review}.
\newblock \textit{Statist. Sci.} 10 273--304.

\bibitem[{Chickering(2002)}]{Chic:2002}
\textsc{Chickering, D.~M.} (2002).
\newblock Learning equivalence classes of {B}ayesian-network structures.
\newblock \textit{J. Mach. Learn. Res.} 2 445--498.

\bibitem[{Consonni \& La~Rocca(2012)}]{Consonni:LaRocca:2012}
\textsc{Consonni, G.} \& \textsc{La~Rocca, L.} (2012).
\newblock Objective {B}ayes factors for {G}aussian directed acyclic graphical
  models.
\newblock \textit{Scand. J. Stat.} 39 743--756.

\bibitem[{Consonni \& Veronese(2008)}]{cons:vero:2008}
\textsc{Consonni, G.} \& \textsc{Veronese, P.} (2008).
\newblock Compatibility of prior specifications across linear models.
\newblock \textit{Statist. Sci.} 23 332--353.

\bibitem[{Cowell et~al.(1999)Cowell, Dawid, Lauritzen \&
  Spiegelhalter}]{Cowe:Dawi:Laur:Spie:1999}
\textsc{Cowell, R.~G.}, \textsc{Dawid, P.~A.}, \textsc{Lauritzen, S.~L.} \&
  \textsc{Spiegelhalter, D.~J.} (1999).
\newblock \textit{Probabilistic Networks and Expert Systems}.
\newblock New York: Springer.

\bibitem[{DasGupta(1996)}]{Dasgupta:1996}
\textsc{DasGupta, A.} (1996).
\newblock 29 review of optimal bayes designs.
\newblock In \textit{Design and Analysis of Experiments}, vol.~13 of
  \textit{Handbook of Statistics}. Elsevier, 1099--1147.

\bibitem[{Dawid(2011)}]{Dawid:2011}
\textsc{Dawid, A.~P.} (2011).
\newblock Posterior model probabilities.
\newblock In P.~S. Bandyopadhyay \& M.~Forster, eds., \textit{Philosophy of
  Statistics}. Elsevier, Amsterdam, 607--630.

\bibitem[{Dawid \& Lauritzen(1993)}]{Dawid:Lauritzen:1993}
\textsc{Dawid, A.~P.} \& \textsc{Lauritzen, S.~L.} (1993).
\newblock Hyper {M}arkov laws in the statistical analysis of decomposable
  graphical models.
\newblock \textit{Ann. Statist.} 21 1272--1317.

\bibitem[{Dawid \& Lauritzen(2001)}]{Dawid:Lauritzen:2001:compatibleprior}
\textsc{Dawid, A.~P.} \& \textsc{Lauritzen, S.~L.} (2001).
\newblock Compatible prior distributions.
\newblock In E.~George, ed., \textit{Bayesian methods with applications to
  science, policy and official statistics: Selected Papers from ISBA 2000, the
  Sixth World Meeting of the International Society for Bayesian Analysis}.
  109--118.

\bibitem[{De~Santis(2004)}]{De:Santis:2004}
\textsc{De~Santis, F.} (2004).
\newblock Statistical evidence and sample size determination for bayesian
  hypothesis testing.
\newblock \textit{J. Statist. Plann. Inference} 124 121--144.

\bibitem[{Drton \& Eichler(2006)}]{Drton:Eichler:2006}
\textsc{Drton, M.} \& \textsc{Eichler, M.} (2006).
\newblock Maximum likelihood estimation in {G}aussian chain graph models under
  the alternative {M}arkov property.
\newblock \textit{Scand. J. Stat.} 33 247--257.

\bibitem[{Eberhardt(2008)}]{Eber:2008}
\textsc{Eberhardt, F.} (2008).
\newblock Almost optimal intervention sets for causal discovery.
\newblock In \textit{Proceedings of the Twenty-Fourth Conference on Uncertainty
  in Artificial Intelligence}, UAI '08. Arlington, Virginia, USA: AUAI Press,
  161--168.

\bibitem[{Etzioni \& Kadane(1993)}]{Etzioni:Kadane:1993}
\textsc{Etzioni, R.} \& \textsc{Kadane, J.~B.} (1993).
\newblock Optimal experimental design for another's analysis.
\newblock \textit{J. Amer. Statist. Assoc.} 88 1404--1411.

\bibitem[{Friedman(2004)}]{Friedman:2004}
\textsc{Friedman, N.} (2004).
\newblock Inferring cellular networks using probabilistic graphical models.
\newblock \textit{Science} 303 799--805.

\bibitem[{Frot et~al.(2019)Frot, Nandy \& Maathuis}]{Frot:et:al:2019}
\textsc{Frot, B.}, \textsc{Nandy, P.} \& \textsc{Maathuis, M.~H.} (2019).
\newblock Robust causal structure learning with some hidden variables.
\newblock \textit{J. R. Stat. Soc. Ser. B. Stat. Methodol.} 81 459--487.

\bibitem[{Geiger \& Heckerman(2002)}]{Geig:Heck:2002}
\textsc{Geiger, D.} \& \textsc{Heckerman, D.} (2002).
\newblock Parameter priors for directed acyclic graphical models and the
  characterization of several probability distributions.
\newblock \textit{Ann. Statist.} 30 1412--1440.

\bibitem[{Gelfand \& Wang(2002)}]{Gelfand:Wang:2002}
\textsc{Gelfand, A.~E.} \& \textsc{Wang, F.} (2002).
\newblock {A simulation-based approach to Bayesian sample size determination
  for performance under a given model and for separating models}.
\newblock \textit{Statist. Sci.} 17 193--208.

\bibitem[{Hauser \& B{\"u}hlmann(2012)}]{Haus:Buhl:2012}
\textsc{Hauser, A.} \& \textsc{B{\"u}hlmann, P.} (2012).
\newblock Characterization and greedy learning of interventional {M}arkov
  equivalence classes of directed acyclic graphs.
\newblock \textit{J. Mach. Learn. Res.} 13 2409--2464.

\bibitem[{Hauser \& B{\"{u}}hlmann(2014)}]{Hauser:Buehlmann:2014}
\textsc{Hauser, A.} \& \textsc{B{\"{u}}hlmann, P.} (2014).
\newblock Two optimal strategies for active learning of causal models from
  interventional data.
\newblock \textit{Int. J. Approx. Reason.} 55 926--939.

\bibitem[{Hauser \& B{\"u}hlmann(2015)}]{Hauser:Buehlm:2015}
\textsc{Hauser, A.} \& \textsc{B{\"u}hlmann, P.} (2015).
\newblock Jointly interventional and observational data: estimation of
  interventional {M}arkov equivalence classes of directed acyclic graphs.
\newblock \textit{J. R. Stat. Soc. Ser. B. Stat. Methodol.} 77 291--318.

\bibitem[{He \& Geng(2008)}]{He:etal:2008}
\textsc{He, Y.} \& \textsc{Geng, Z.} (2008).
\newblock Active learning of causal networks with intervention experiments and
  optimal designs.
\newblock \textit{J. Mach. Learn. Res.} 9 2523--2547.

\bibitem[{He et~al.(2013)He, Jia \& Yu}]{He:etal:2013}
\textsc{He, Y.}, \textsc{Jia, J.} \& \textsc{Yu, B.} (2013).
\newblock Reversible {MCMC} on {M}arkov equivalence classes of sparse directed
  acyclic graphs.
\newblock \textit{Ann. Statist.} 41 1742--1779.

\bibitem[{Hyttinen et~al.(2013)Hyttinen, Eberhardt \&
  Hoyer}]{Hytt:etal:JMLR:2013}
\textsc{Hyttinen, A.}, \textsc{Eberhardt, F.} \& \textsc{Hoyer, P.~O.} (2013).
\newblock Experiment selection for causal discovery.
\newblock \textit{J. Mach. Learn. Res.} 14 3041--3071.

\bibitem[{Imbens(2020)}]{Imbens:2020}
\textsc{Imbens, G.~W.} (2020).
\newblock Potential outcome and directed acyclic graph approaches to causality:
  relevance for empirical practice in economics.
\newblock \textit{J. Econ. Lit.} 58 1129--1179.

\bibitem[{Jeffreys(1961)}]{Jeffreys:1961}
\textsc{Jeffreys, H.} (1961).
\newblock \textit{Theory of Probability (3rd Edition)}.
\newblock Oxford, University Press.

\bibitem[{Johnson \& Rossell(2010)}]{Johnson:Rossell:2010}
\textsc{Johnson, V.~E.} \& \textsc{Rossell, D.} (2010).
\newblock On the use of non-local prior densities in {B}ayesian hypothesis
  tests.
\newblock \textit{J. R. Stat. Soc. Ser. B. Stat. Methodol.} 72 143--170.

\bibitem[{Kalisch \& B{{\"u}}hlmann(2007)}]{Kalish:Buehlmann:2007}
\textsc{Kalisch, M.} \& \textsc{B{{\"u}}hlmann, P.} (2007).
\newblock Estimating high-dimensional directed acyclic graphs with the
  pc-algorithm.
\newblock \textit{J. Mach. Learn. Res.} 8 613--636.

\bibitem[{Kass \& Raftery(1995)}]{Kass:Raftery:1995}
\textsc{Kass, R.~E.} \& \textsc{Raftery, A.~E.} (1995).
\newblock Bayes factors.
\newblock \textit{J. Amer. Statist. Assoc.} 90 773--795.

\bibitem[{Koller \& Friedman(2009)}]{Koller:Friedman:2009}
\textsc{Koller, D.} \& \textsc{Friedman, N.} (2009).
\newblock \textit{Probabilistic Graphical Models: Principles and Techniques}.
\newblock Adaptive computation and machine learning. MIT Press.

\bibitem[{Lauritzen(1996)}]{Laur:1996}
\textsc{Lauritzen, S.~L.} (1996).
\newblock \textit{Graphical Models}.
\newblock Oxford University Press.

\bibitem[{Lindley(1972)}]{lind:1972}
\textsc{Lindley, D.~V.} (1972).
\newblock \textit{1. Bayesian Statistics, a Review}.
\newblock 1--74.

\bibitem[{Lindley(1997)}]{Lindley:1997}
\textsc{Lindley, D.~V.} (1997).
\newblock The choice of sample size.
\newblock \textit{J. R. Stat. Soc. D. Stat.} 46 129--138.

\bibitem[{Maathuis et~al.(2009)Maathuis, Kalisch \& Bühlmann}]{Maat:etal:2009}
\textsc{Maathuis, M.~H.}, \textsc{Kalisch, M.} \& \textsc{Bühlmann, P.}
  (2009).
\newblock {Estimating high-dimensional intervention effects from observational
  data}.
\newblock \textit{Ann. Statist.} 37 3133--3164.

\bibitem[{Madigan et~al.(1996)Madigan, Andersson, Perlman \&
  Volinsky}]{Madigan:et:al:1996}
\textsc{Madigan, D.}, \textsc{Andersson, S.~A.}, \textsc{Perlman, M.~D.} \&
  \textsc{Volinsky, C.~T.} (1996).
\newblock Bayesian model averaging and model selection for {M}arkov equivalence
  classes of acyclic digraphs.
\newblock \textit{Commun. Stat. - Theor. M.} 25 2493--2519.

\bibitem[{Meganck et~al.(2006)Meganck, Leray \& Manderick}]{Mega:etal:2006}
\textsc{Meganck, S.}, \textsc{Leray, P.} \& \textsc{Manderick, B.} (2006).
\newblock Learning causal bayesian networks from observations and experiments:
  A decision theoretic approach.
\newblock In V.~Torra, Y.~Narukawa, A.~Valls \& J.~Domingo-Ferrer, eds.,
  \textit{Modeling Decisions for Artificial Intelligence}. Berlin, Heidelberg:
  Springer Berlin Heidelberg, 58--69.

\bibitem[{Muirhead(1982)}]{Muirhead:1982}
\textsc{Muirhead, R.} (1982).
\newblock \textit{Aspects of Multivariate Statistical Theory}.
\newblock John Wiley \& Sons, Ltd.

\bibitem[{Nagarajan et~al.(2013)Nagarajan, Scutari \& Lèbre}]{Nagarajan:2013}
\textsc{Nagarajan, R.}, \textsc{Scutari, M.} \& \textsc{Lèbre, S.} (2013).
\newblock \textit{Bayesian Networks in R: With Applications in Systems
  Biology}.
\newblock Springer Publishing Company, Incorporated.

\bibitem[{O'Hagan(1995)}]{Ohagan:1995}
\textsc{O'Hagan, A.} (1995).
\newblock Fractional {B}ayes factors for model comparison.
\newblock \textit{J. R. Stat. Soc. Ser. B. Stat. Methodol.} 57 99--138.

\bibitem[{O'Hagan \& Stevens(2001)}]{O'Hagan:Stevens:2001}
\textsc{O'Hagan, A.} \& \textsc{Stevens, J.} (2001).
\newblock Bayesian assessment of sample size for clinical trials of
  cost-effectiveness.
\newblock \textit{Med. Decis. Making} 21 219--30.

\bibitem[{Pan \& Banerjee(2021)}]{Pan:Banerjee:2021}
\textsc{Pan, J.} \& \textsc{Banerjee, S.} (2021).
\newblock A unifying bayesian approach for sample size determination using
  design and analysis priors.
\newblock \textit{arXiv preprint arXiv:2112.03509} .

\bibitem[{Pearl(2000)}]{Pear:2000}
\textsc{Pearl, J.} (2000).
\newblock \textit{{Causality: Models, Reasoning, and Inference}}.
\newblock Cambridge University Press, Cambridge.

\bibitem[{Pearl(2003)}]{Pearl:review:Test:2003}
\textsc{Pearl, J.} (2003).
\newblock {Statistics and causal inference: A review}.
\newblock \textit{Test}  281--345.

\bibitem[{Peng et~al.(2020)Peng, Shen \& Pan}]{Peng:et:al:2020}
\textsc{Peng, S.}, \textsc{Shen, X.} \& \textsc{Pan, W.} (2020).
\newblock {Reconstruction of a directed acyclic graph with intervention}.
\newblock \textit{Electron. J. Stat.} 14 4133--4164.

\bibitem[{Peters \& Bühlmann(2013)}]{Peters:Buehlmann:2014}
\textsc{Peters, J.} \& \textsc{Bühlmann, P.} (2013).
\newblock {Identifiability of {G}aussian structural equation models with equal
  error variances}.
\newblock \textit{Biometrika} 101 219--228.

\bibitem[{Press(1982)}]{Press:1982}
\textsc{Press, S.~J.} (1982).
\newblock \textit{Applied multivariate analysis: using Bayesian and frequentist
  methods of inference}.
\newblock Krieger Publishing Company, Malabar, FL.

\bibitem[{{R Core Team}(2021)}]{R:core:team}
\textsc{{R Core Team}} (2021).
\newblock \textit{R: A Language and Environment for Statistical Computing}.
\newblock R Foundation for Statistical Computing, Vienna, Austria.

\bibitem[{Raiffa \& Schlaifer(1961)}]{Raiffa:Schlaifer:1961}
\textsc{Raiffa, H.} \& \textsc{Schlaifer, R.} (1961).
\newblock \textit{Applied Statistical Decision Theory}.
\newblock Harvard Business School Publications. Division of Research, Graduate
  School of Business Adminitration, Harvard University.

\bibitem[{Royall(1997)}]{Royal:1997}
\textsc{Royall, R.} (1997).
\newblock \textit{Statistical Evidence: A likelihood paradigm (1st ed.)}.
\newblock Routledge.

\bibitem[{Royall(2000)}]{Royall:2000}
\textsc{Royall, R.} (2000).
\newblock On the probability of observing misleading statistical evidence.
\newblock \textit{J. Amer. Statist. Assoc.} 95 760--768.

\bibitem[{Sachs et~al.(2005)Sachs, Perez, Pe’er, Lauffenburger \&
  Nolan}]{Sachs:etal:2005}
\textsc{Sachs, K.}, \textsc{Perez, O.}, \textsc{Pe’er, D.},
  \textsc{Lauffenburger, D.} \& \textsc{Nolan, G.} (2005).
\newblock Causal protein-signaling networks derived from multiparameter
  single-cell data.
\newblock \textit{Science} 308 523--529.

\bibitem[{Sadeghi(2017)}]{Sadeghi:2017}
\textsc{Sadeghi, K.} (2017).
\newblock Faithfulness of probability distributions and graphs.
\newblock \textit{J. Mach. Learn. Res.} 18 1--29.

\bibitem[{Sch\"{o}nbrodt \& Wagenmakers(2017)}]{Schonbrodt:Wagen:2017}
\textsc{Sch\"{o}nbrodt, F.~D.} \& \textsc{Wagenmakers, E.} (2017).
\newblock Bayes factor design analysis: Planning for compelling evidence.
\newblock \textit{Psychon. B. Rev.} 25 128--142.

\bibitem[{Shojaie \& Michailidis(2009)}]{Shoj:Mich:2009}
\textsc{Shojaie, A.} \& \textsc{Michailidis, G.} (2009).
\newblock Analysis of gene sets based on the underlying regulatory network.
\newblock \textit{J. Comput. Biol.} 16 407--26.

\bibitem[{Sonntag et~al.(2015)Sonntag, Peña \& Gómez-Olmedo}]{Sonn:etal:2015}
\textsc{Sonntag, D.}, \textsc{Peña, J.~M.} \& \textsc{Gómez-Olmedo, M.}
  (2015).
\newblock Approximate counting of graphical models via {MCMC} revisited.
\newblock \textit{Int. J. Intell. Syst.} 30 384--420.

\bibitem[{Spiegelhalter et~al.(2003)Spiegelhalter, Abrams \&
  Myles}]{Spiegelhalter:et:al:book:2003}
\textsc{Spiegelhalter, D.}, \textsc{Abrams, K.} \& \textsc{Myles, J.} (2003).
\newblock \textit{Bayesian Approaches to Clinical Trials and Health‐Care
  Evaluation}.
\newblock John Wiley \& Sons, Ltd.

\bibitem[{Spiegelhalter \& Freedman(1986)}]{Spiegelhalter:Freedman:1986}
\textsc{Spiegelhalter, D.~J.} \& \textsc{Freedman, L.~S.} (1986).
\newblock A predictive approach to selecting the size of a clinical trial,
  based on subjective clinical opinion.
\newblock \textit{Stat. Med.} 5 1--13.

\bibitem[{Spirtes et~al.(2000)Spirtes, Glymour \&
  Scheines}]{Spir:Glym:Sche:2000}
\textsc{Spirtes, P.}, \textsc{Glymour, C.} \& \textsc{Scheines, R.} (2000).
\newblock \textit{{Causation, Prediction and Search (2nd edition)}}.
\newblock Cambridge, MA: The MIT Press.

\bibitem[{Squires et~al.(2020)Squires, Magliacane, Greenewald, Katz, Kocaoglu
  \& Shanmugam}]{squi:etal:arxiv:2020}
\textsc{Squires, C.}, \textsc{Magliacane, S.}, \textsc{Greenewald, K.},
  \textsc{Katz, D.}, \textsc{Kocaoglu, M.} \& \textsc{Shanmugam, K.} (2020).
\newblock Active structure learning of causal {DAG}s via directed clique trees.
\newblock In \textit{Proceedings of the 34th International Conference on Neural
  Information Processing Systems}, NIPS '20. Red Hook, NY, USA: Curran
  Associates Inc.

\bibitem[{Stefan et~al.(2022)Stefan, Sch\"{o}nbrodt, Evans \&
  Wagenmakers}]{Stefan:et:al:2022}
\textsc{Stefan, A.~M.}, \textsc{Sch\"{o}nbrodt, F.~D.}, \textsc{Evans, N.~J.}
  \& \textsc{Wagenmakers, E.~J.} (2022).
\newblock {Efficiency in sequential testing: Comparing the sequential
  probability ratio test and the sequential Bayes factor test}.
\newblock \textit{Behav. Res. Methods} 54 1554--3528.

\bibitem[{Tong \& Koller(2001)}]{Tong;Koll:2001}
\textsc{Tong, S.} \& \textsc{Koller, D.} (2001).
\newblock Active learning for structure in bayesian networks.
\newblock In \textit{Proceedings of the 17th International Joint Conference on
  Artificial Intelligence - Volume 2}, IJCAI '01. San Francisco, CA, USA:
  Morgan Kaufmann Publishers Inc., 863--869.

\bibitem[{Verma \& Pearl(1990)}]{Verma:Pearl:1990}
\textsc{Verma, T.} \& \textsc{Pearl, J.} (1990).
\newblock Equivalence and synthesis of causal models.
\newblock In \textit{Proceedings of the Sixth Annual Conference on Uncertainty
  in Artificial Intelligence}, UAI 90. New York, NY, USA: Elsevier Science
  Inc., 255--270.

\bibitem[{von K{\"u}gelgen et~al.(2019)von K{\"u}gelgen, Rubenstein,
  Sch{\"o}lkopf \& Weller}]{Klugel:et:al:2019}
\textsc{von K{\"u}gelgen, J.}, \textsc{Rubenstein, P.~K.},
  \textsc{Sch{\"o}lkopf, B.} \& \textsc{Weller, A.} (2019).
\newblock Optimal experimental design via bayesian optimization: active causal
  structure learning for gaussian process networks.
\newblock In \textit{NeurIPS 2019 Workshop Do the right thing: machine learning
  and causal inference for improved decision making}.

\bibitem[{Weiss(1997)}]{Weiss:1997}
\textsc{Weiss, R.} (1997).
\newblock Bayesian sample size calculations for hypothesis testing.
\newblock \textit{J. R. Stat. Soc. D. Stat.} 46 185--191.

\bibitem[{Yang et~al.(2018)Yang, Katcoff \& Uhler}]{Yang:Uhler:2018}
\textsc{Yang, K.}, \textsc{Katcoff, A.} \& \textsc{Uhler, C.} (2018).
\newblock Characterizing and learning equivalence classes of causal {DAG}s
  under interventions.
\newblock In J.~Dy \& A.~Krause, eds., \textit{Proceedings of the 35th
  International Conference on Machine Learning}, vol.~80 of \textit{Proceedings
  of Machine Learning Research}. PMLR, 5541--5550.

\end{thebibliography}

\end{document}